# Enhancement in Reliability for Multi-core system consisting of One Instruction Cores

Shashikiran Venkatesha [#, a], Ranjani Parthasarathi [b]

[a] Research scholar, Department of Information Science and Technology, College of Engineering Guindy Anna University Chennai 600025 Tamil Nadu India.

[b] Professor, Department of Information Science and Technology, College of Engineering Guindy Anna University Chennai 600025 Tamil Nadu India.

Abstract

Rapid CMOS device size reduction resulted in billions of transistors on a chip have led to integration of many cores leading to many challenges such as increased power dissipation, thermal dissipation, occurrence of transient faults and permanent faults. The mitigation of transient faults and permanent faults at the core level has become an important design parameter in a multi-core scenario. Core level techniques is a redundancy-based fault mitigation technique that improves the lifetime reliability of multi-core systems. In an asymmetric multi-core system, the smaller cores provide fault tolerance to larger cores is a core level fault mitigation technique that has gained momentum and focus from a large number of researchers. The paper presents an economical, asymmetric multi-core system with one instruction cores (MCS-OIC). The term "Hardware Cost Estimation" signifies power and area estimation for MCS-OIC. In MCS-OIC, OIC is a warm standby redundant core. OICs provide functional support to conventional cores for shorter periods of time. To evaluate the idea, different configurations of MCS-OIC is synthesized using FPGA and ASIC. The maximum power overhead and maximum area overhead are 0.46% and 11.4% respectively. The behaviour of OICs in MCS-OIC is modelled using a One-Shot System (OSS) model for reliability analysis. The model parameters namely, readiness, wakeup probability and start-up-strategy for OSS are mapped to the multi- core systems with OICs. Expressions for system reliability is derived. System reliability is estimated for special cases.

## 1. Introduction

The number of transistors on a chip has been doubling every 18 months as predicted by Moore's law [1,2]. The performance scaling in a single core has followed the trajectory laid down by Moore's law. Transistor density has been increasing by about 35% per year and the die size has been increasing by 10% to 20% per year [3]. Despite the gains, performance per watt/per mm die area has been diminishing. As the size of the transistor is reduced, the power density increases thereby creating a "power wall" that has limited the processor frequency at 4 GHz in 2006 [4]. To tackle the power wall, chip designers and fabricators started focussing on the multi-core processors [5] extended with smaller specialized processors. Since then, the trend has been on increasing the number of cores, which in turn increases the number of transistors on a chip. This rapid integration with high transistor density, has also increased the vulnerability of cores [6]. Quicker CMOS integration has resulted in challenges including heightened power/thermal dissipation, occurrence of faults in the circuits, and lack of reliability [7]. This has resulted in the need for fault mitigation and fault tolerance techniques. This paper presents a fault tolerance solution in a multi-core scenario.

Faults can be broadly classified into three types: permanent faults or hard faults, intermittent faults and transient faults [8]. Electro migration, stress migration, thermal cycling, time-dependent dielectric breakdown, and negative bias temperature instability cause permanent faults. Intermittent faults are caused by oxide degradation, manufacturing residuals and process variations. Intermittent faults occur irregularly and frequently for a period of time. Transient faults or soft errors are caused by cosmic rays, alpha particles, air pollution, humidity, noise and erratic voltage fluctuations [9]. Researchers have predicted about an eight percent increase in soft-error rate per logic state bit in each technology generation [10]. According to the International Roadmap for Device and System (IRDS) roadmap 2017, device scaling will touch the physical limits with failures reaching 1 failure per hour. It can be seen, from the Figure 1.1 that, at 16nm process node size, a chip with 100 cores could come across one failure every hour due to soft errors [11,12].

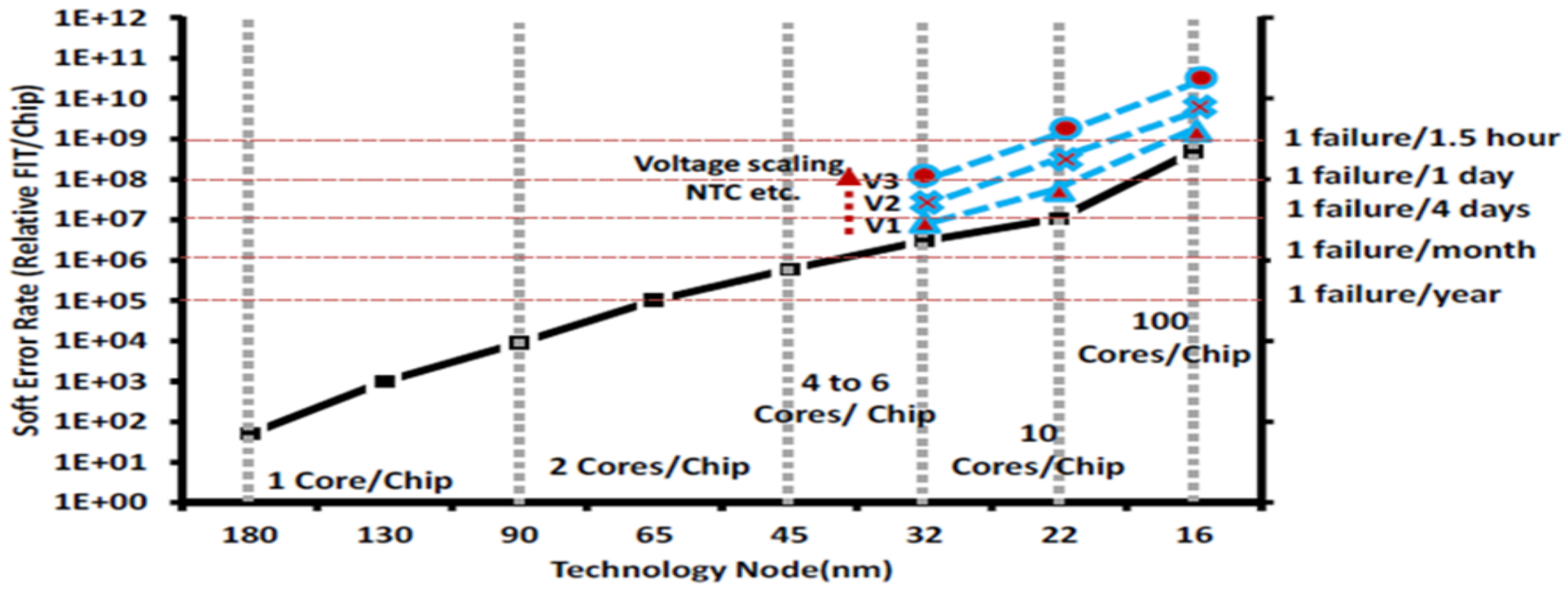

(Source : Shivakumar et al. 2002[11]).

**Figure 1.1 SER rates at various technology node**

As seen in Figure 1.1, the total Failure in Time (FIT) per chip increases with number of cores per chip increasing. In order to accommodate higher number of cores per chip, (1) total FIT per chip has to be maintained constant (or no change), and (2) soft error rate per core needs to be reduced. In the present day processor cores, the frontend of the core comprises of decode queue, instruction translation lookaside buffer, and latches. The backend of the core comprises of arithmetic logic unit, register files, data translation lookaside buffer, reorder buffers, memory order buffer, and issue queue. Soft error rate from backend and the frontend of the core is 74.48% and 25.22% respectively. In the present processor cores, latches are hardened [13,14], cache and large memory arrays are protected using error correcting codes (ECC) [15]. The SER from backend of the processor is more when compared to front end and is mainly due to arithmetic logic unit.

The FIT from the arithmetic logic unit of the processor core has started reaching higher levels which needs robust fault mitigation approaches for present and future processors. Hence addressing the reliability issues of the core (arithmetic logic unit in backend) is more significant in improving the reliability of the multi-core system [16,17]. Conventional approaches to handle soft errors consumes more power and area. Hence, this paper presents heterogeneous model with "low cost" (less power and lesser area) fault tolerant cores to improve reliability of multi-core systems.

We introduce heterogeneous multi-core system in which conventional cores are supported by One Instruction Cores (OIC) [18]. A 32-bit OIC is designed to provide fault tolerance to a multi-core system with 32-bit integer instructions of conventional MIPS cores. OIC executes only one instruction, namely, "subleq – subtract if less than or equal". The OIC has three conventional subtractors and an additional self-checking subtractor. A conventional core that detects faults in one of the functional units (i.e., ALU) sends the opcode with operands to the OIC. The OIC emulates the instruction by repetitively executing the subleq instruction in a predetermined manner. Our earlier work focused on area, power and reliability analysis of OIC only. In this paper, we present an integrated solution i.e., area, power and reliability analysis for multi-core system consisting conventional cores (MIPS core(s)) supported by OICs, a part of our earlier work [45].

OIC is a standby redundancy core that provides functional support to conventional cores. From Figure 1.1, it is observed that cores would encounter one soft error per year at 90nm which goes up to one soft error per hour at 16nm. This indicates that it is sufficient OICs support to conventional cores intermittently. Hence, OIC support for conventional cores is needed for short periods of time. The reliability model presented in this paper captures the computational support provided by OIC to ALU of the conventional cores for shorter intervals of time. One Shot System (OSS) which is one such model, is adopted for modelling the multi-core system with OICs. In OSS, the system is expected to perform one or more tasks only once. It could be repeated but independent of prior task execution. In this paper, we present OSS based reliability model, model parameters and estimation of system reliability for Multi-Core System using OICs (shortly as MCS-OIC).

In a MCS-OIC, when many conventional cores and OICs exist, choice of OIC and their functions that are required by the conventional cores becomes an important consideration that determines the reliability of the system. This choice of OICs and their functions is viewed as a redundancy allocation problem (RAP). Further, another factor to be considered is that the reliability of migrating instructions from conventional core to OICs and executing the function on OICs, *r(t)*, is imprecise. This impreciseness could be due to manufacturing complexity, or complexity in design and the environment where it is deployed. In order to address impreciseness, conventionally three approaches are available, namely, interval analysis, fuzzy and stochastic fuzzy. Among these, Interval analysis is used to address impreciseness. Hence Interval valued reliabilities are assigned to *r(t)*. Thus, the RAP is referred to as Interval RAP. In this paper, we present formulation for Interval RAP and solution using a modified Genetic Algorithm (GA) and Particle Swarm Optimization (PSO) that addresses the interval numbers, part of our earlier [46].

We summarize the contributions and enumerate the same below.

(a) Power and Area analysis for MCS-OIC.

(b) OSS based Reliability model used to estimate the system reliability for the MCS-OIC.

(c) Formulation of interval RAP and solution for system reliability using proposed GA and PSO for interval numbers.

The remaining sections of this paper are organized as follows: related work on reliability modelling for standby redundancy systems in multi-core scenario are presented in the section titled “2. Background on Reliability modelling for standby redundancy systems”; Arithmetic and Order relations for interval numbers are presented in the section titled “3. Background on Interval Analysis”; Microarchitecture of MCS-OIC (consisting of One MIPS core and One OIC), simulation results for power and area are presented in the section titles “4. Micro-architecture of MCS-OIC and Experimental results”; OSS model, model parameters, different configurations of MCS–OIC and mapping OSS models parameters for MCS-OIC are presented in the section titled “5. OSS and Model parameters for MCS-OIC”; Reliability modelling assumptions, Notations and deriving an expression for system reliability for MCS-OIC are presented in the section titled “6. System reliability for MCS-OIC”; discussion on RAP, Interval RAP and its inherent properties are presented in the section titled “7. Interval RAP and Its Inherent Characteristics”; solution for Interval RAP using proposed GA (with sensitivity analysis) and PSO are presented in the sections titled “8. Genetic Algorithm for Interval RAP” and “9. Particle swarm optimization for Interval RAP” respectively; and the conclusion of this paper is presented in the section titled “10. Conclusion”.

## 2. Background on Reliability modelling for standby redundancy systems

Traditionally, classical two state series and parallel systems have standby redundancy components with the ability to replace faulty components. Two state system can be in two states: success and failed state. Two state k-out-of-n systems are popular in abstracting the standby redundancy-based fault-tolerant series/parallel systems. A system consisting of n components that works (or “is good”) if and only if at least k of the n components work is called k-out-of-n systems: G system, as introduced by Birnbaum et al. [19]. A system of n components that fails if and only if at least k components of n fail is called k-out-of-n systems: F system. The Multistate is a k-out-of-n systems: G system, the reliability R (k, n, and j) is the probability that k components are in the high performing state (j) or above. Similarly, if at least k components are below low performing state (l), such n-component system is called k-out-of-n systems: F system.

Seminal contributions on reliability modelling for hot standby system [20] discuss mean residual life time of k-out-of-n systems (both in series and parallel structures). The failure time in exponential distribution [21], Normal distribution [22] is used to analyse the standby redundancy system. Gnedenko et al. [23] studied 1-out-of-n warm standby systems with identical components using general failure time distributions based on the concept of equivalent age. Li et al. [24] studied hazard rates of components of warm standby systems. K-out-of-n systems warm standby system's reliability characteristics are discussed by [25]. Huang & Xu [26] proposed three schemes for homogeneous many-core systems and they are (a) Standby Redundant System (k-out-of-n systems: G system), (b) Gracefully Degrading System (GDS) (all n cores are active cores), and (c) processor rotation scheme. This approach uses Weibull distribution to estimate the system reliability. Extension of the system lifetime is reported in the experimental results with the increase in number of cores.

In all the methods discussed above, the k cores are spare cores and are used when active cores fail. The OICs in MCS-OIC provide functional redundancy to the other partially active conventional cores for a smaller interval of time. The component level redundancy and functional level redundancy are the key characteristics of MCS-OIC and these aspects cannot be modelled using traditional k-out-of-n systems. Therefore, k-out-of-n systems based reliability modelling is not a suitable model for multi-core systems with OICs.

As explained above, One Shot System (OSS) is one such model that captures the functional support provided by OICs in MCS-OIC. OSS have been used to model nuclear weapons systems [27], Ad-hoc sensor networks [28], and missile systems [29]. OSS model used for Ad-hoc sensor network is modified entirely and adapted to MCS-OIC. In MCS-OIC, Reasonably, under two circumstances One Shot System (OSS) correctly abstracts the functional redundancy provided by OICs. They are as follows:

(a) When the cores are struck by soft errors, OICs provides a functional support for the conventional cores, while the core recovers to a stable state later.

(b) Execution of instructions which has high failure probability on conventional core can be migrated to OICs. Only those instructions are executed on OICs.

OSS modelling of MCS –OIC is presented in the Section 5.

## 3. Background on Interval Analysis

Interval analysis was first introduced by [30] to address impreciseness. Interval arithmetic and Order relations are important for crossover/mutation and selection process for finding the better solution in the Genetic algorithm. In PSO, Order relations is used in updating $G_{best}$, $L_{best,i}$ and $P_{best,i}$ parameters and Interval arithmetic in updating velocities. Interval arithmetic and Order relations are discussed below. Interval analysis is used to address impreciseness in electric power system [31], microelectronics circuits [32], structure safety analysis [33], and damage identification [34]. Wang et al. [35] proposed interval analysis for standby redundancy-based system optimization problems. Bhunia & Samanta [36] proposed interval analysis for theoretical multi-objective optimization problem in the interval domain. Gupta & Bhunia [37], used interval-valued reliability for components in the series systems.

### 3.1. Arithmetic for interval numbers

Consider W to be an interval valued number. It is defined as set of real numbers '$\boldsymbol{a}$', such that, $w_L \leq a \leq w_R$ , $i.e.$, $a \in [w_L, w_R], w_L, w_R \in R$. $R$ is set of real numbers. $w_L$ and $w_R$ are left and right boundaries of the interval respectively. W is expressed in Equation (1).

$$W = [w_L, w_R] = \{ a : w_L \leq a \leq w_R , a \in R \} \quad (1)$$

An interval valued number W can also be expressed in terms of radius and centre and is expressed in Equation (2) 6.2.

$$W = < w_c, w_r > = \{ a : w_c - w_r \leq a \leq w_c + w_r, a \in R \} \quad (2)$$

Where $w_c = (w_L + w_R)/2$ and $w_r = (w_R - w_L)/2$ are the centre and the radius of the interval, respectively, and R is the set of real numbers.

Definition 1.1: Arithmetic operations: Consider $X = [X_L, X_R]$ and $Y = [Y_L, Y_R]$ are closed intervals. Then addition, subtraction multiplication and division are defined as given below.

Addition of two intervals X and Y is given in Equation (3).

$$X + Y = [X_L + Y_L, X_R + Y_R] \quad (3)$$

Subtraction of two intervals and Y is given in Equation (4).

$$X - Y = [X_L - Y_L, X_R - Y_R] \quad (4)$$

Multiplication of an interval by a real number λ is given in Equations (5) and (6).

$$\lambda.X = \lambda.[X_L, X_R] = [\lambda X_L, \lambda X_R] \;\; if \; \lambda \geq 0 \quad (5)$$

$$\lambda.X = \lambda.[X_L, X_R] = [\lambda X_R, \lambda X_L] \;\; if \; \lambda < 0 \quad (6)$$

Multiplication of two intervals is given in the Equation (7).

$$X \text{ x } Y = [\min(X_L Y_L, X_L Y_R, X_R Y_L, X_R Y_R), \max(X_L Y_L, X_L Y_R, X_R Y_L, X_R Y_R)] \quad (7)$$

Division of interval Y by interval X is given in the Equation (8).

$$\frac{Y}{X} = Y \text{ x } \frac{1}{X} = [Y_L, Y_R] \text{ x } [\frac{1}{X_L}, \frac{1}{X_R}] \text{ provided } X_L \neq 0, X_R \neq 0. \quad (8)$$

Definition 1.2**:** Power of interval: Consider X =$[X_L, X_R]$. For a non-negative integer N, the N$^{th}$ power of the interval X is given in Equation (9).

$$X^N = [1,1], \qquad if \; N = 0.$$

$$X^N = [X_L^N, X_R^N], if \; X_L \geq 0, or \; if \; N \; is \; odd$$

$$X^N = [X_R^N, X_L^N], if \; X_R \leq 0, or \; if \; N \; is \; even$$

$$X^N = [0, \max(X_L^N, X_R^N)], if \; X_L \leq 0 \leq X_R, N > 0 \; is \; even. \quad (9)$$

3.2. Order relations for interval numbers

The order relations between pairs of intervals valued numbers rest on the perception the decision taker has in the optimization problem. The selection of the best alternative between pair of intervals valued numbers for maximization problems is discussed in this section.

Two interval numbers X= $[X_L, X_R]$ and Y = $[Y_L, Y_R]$ can be categorized into the following three types.

(i) Both the intervals are disjoint

(ii) Partially overlapping intervals

(iii) One interval contained in the other.

Mahato & Bhunia [38] developed definitions for these cases in the perspective of optimistic and pessimistic decision maker's standpoint on order relations for maximization problems.

***Optimistic decision making:***

Definition 1.3**.** The order relation $\geq_{o\,max}$ between X and Y for maximization problems is given in Inequality 6.10 and (11).

$$X \geq_{o\,max} Y \; iff \; X_R \geq Y_R \tag{10}$$

$$X >_{o\,max} Y \; iff \; X \geq_{o\,max} Y \; and \; X \neq Y. \tag{11}$$

The optimistic decision taker accepts the interval X which is superior to Y. Order relation $\geq_{o\,max}$ is symmetric but not transitive.

***Pessimistic decision-making:***

Definition 1.4**.** The order relation $\geq_{o\,max}$ between the intervals X = $[X_L, X_R] = \langle X_c, X_r \rangle$ and Y = $[Y_L, Y_R] = \langle Y_c, Y_r \rangle$ from pessimistic decision taker’s standpoint for maximization problems is given in Inequality (12) and (13) .

$$X >_{p\,max} Y \; iff \; X_c > Y_c \; for \text{ disjoint } and \; partially \; ovelapped \; intervals. \tag{12}$$

$$X >_{p\,max} Y \; iff \; X_c \geq Y_c \; and \; X_r < Y_r \; when \; one \; interval \; is \; contained \; in \; the \; other. \tag{13}$$

(6.13)

Optimistic decision making may have to be considered for $X_c > Y_c \; and \; X_r > Y_r$.

## 4. Micro-architecture of MCS-OIC and Experimental results

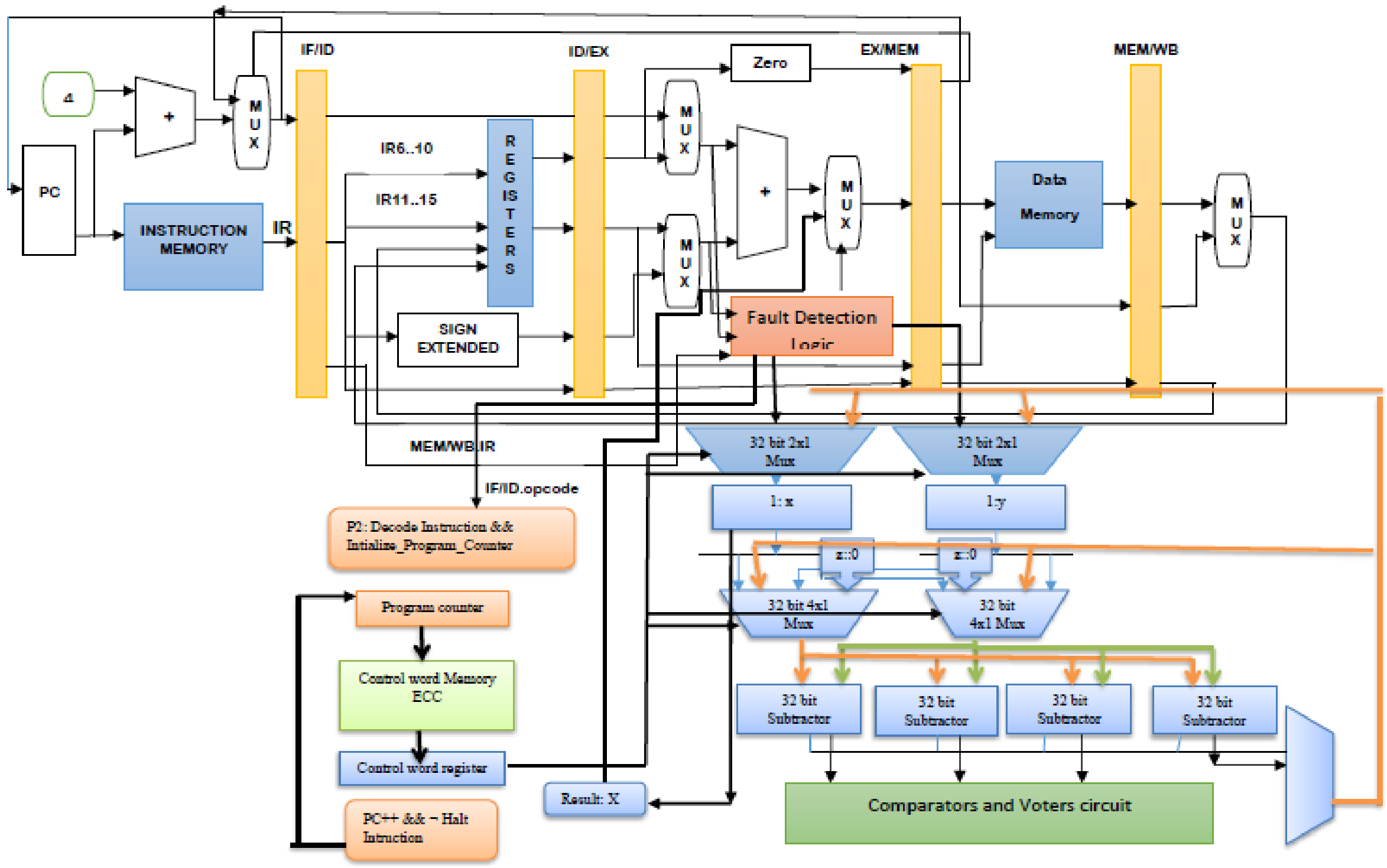

**Figure 1.2 Micro-architecture of multi-core system consisting of one 32-bit MIPS core and one 32-bit OIC.**

In this section, Microarchitecture of MCS-OIC (consisting of One MIPS core and One OIC), simulation results for power and area are presented. A Multicore system comprising one 32 bit MIPS core and one 32bit OIC occupying the upper half and lower half portions respectively in the micro-architecture, is shown in the Figure 1.2. The MIPS core is a five-stage pipelined scalar processor. Instruction Fetch (IF), Instruction Decode (ID), Execution (EXE), Memory access (MEM) and Write Back (WB) are the five stages in the MIPS pipeline. IF/ID, ID/EXE, EXE/MEM, and MEM/WB are the pipeline registers. PC is a program counter and LMD, Imm, A, B, IR, NPC, Aluoutput, and Cond are temporary registers that hold state values between clock cycles of one instruction. The Fault detection logic (FDL) detects faults in all the arithmetic instructions (except logical instructions) by concurrently executing the instructions. The results of ID/EXE.Aluoutput and FDL are compared to detect the fault. If a fault is found then the pipeline is stalled. The IF/ID.opcode (in IR) and operands ID/EXE.A and ID/EXE.B are transferred to OIC. The IF/ID.opcode is decoded and concurrently ID/EXE.A and ID/EXE.B values are loaded into the OIC registers (X & Y). The OIC.PC is initialized and simultaneously first control word from memory is loaded into its control word register. During every clock cycle, the control bits from control word register are sent to the selection lines of the multiplexer that control the input lines to the subtractors. At every clock cycle, subtraction is performed to emulate the instruction sent from the MIPS core. Finally, the computed result is loaded into MEM/WB.Aluoutput and the MIPS pipeline operation is resumed.

The area and power for the micro-architecture of multi-core system (one MIPS core with one OIC) shown in the Figure 1.2, are estimated using ASIC simulation. FPGA synthesis is used to verify the prototype (MCS-OIC) functioning and determine the number of logical elements in the circuit. The details of simulation set-up and parameters estimated are given in the Table 1. The multi-core system occupies a total area of 306283 µm2 and consumes a total power of 1.1554W. The FDL occupies an area of 6203 µm2 which is 2% of the total area occupied by the system. The OIC occupies an area of 8122 µm2 which is 2.6 % of the total area occupied by the system. The FDL consumes a power of 1.2mW and OIC consumes a power of 1.4mW which are negligible when compared to the total power. For various configurations of multicore systems, system power, system area, additional power (due to FDL and OIC), and additional area (due to FDL and OIC) are tabulated in Table 2.

Number of logical elements per OIC and per MIPS core are tabulated in Table 3. Logical elements per OIC is just 2.65% of one MIPS core and 2.58% of MCS-OIC.

**Table 1 Simulation set-up and Parameter Evaluation - Usage**

| Platform | Experimental setup | Parameter Evaluation - Usage |
|---|---|---|
| FPGA | Quartus prime Cyclone IVE with device EP4CE115FE29C7 | (1) Number of Logical elements and Register count for OIC and MIPS core are evaluated. Usage: Logical elements count used in reliability analysis presented in Section VI |
| ASIC | Cadence Encounter (R) RTL Compiler RC14.28 –V14.20 (Cadence design systems 2004) TSMC 90nm technology library (tcbn90lphptc 150). | (1) Area and Power for MCS-OIC |

**Table 2 Power and Area estimation for different configurations of multi-core systems using ASIC simulation**

| Configurations | System power(W) | System area (μm$^2$) | Additional power | | Additional area | |
|---|---|---|---|---|---|---|
| | | | (mW) | % of system power | (μm2) | % of System area |
| one MIPS core + one OIC | 1.1554 | 306283 | 2.6 | 0.2 | 14325 | 4.6 |
| one MIPS core + two OICs | 1.1568 | 314405 | 4 | 0.34 | 22447 | 7.1 |
| one MIPS core + four OICs | 1.1586 | 338771 | 5.4 | 0.46 | 38691 | 11.4 |
| two MIPS core + one OIC | 2.3098 | 604444 | 3.8 | 0.16 | 20528 | 3.3 |
| two MIPS core + two OICs | 2.3112 | 612566 | 5.2 | 0.22 | 28650 | 4.6 |
| two MIPS core + four OICs | 2.3140 | 628810 | 8 | 0.34 | 44894 | 7.13 |
| four MIPS core + one OIC | 4.6222 | 1175954 | 6.2 | 0.13 | 32934 | 2.8 |
| four MIPS core + two OICs | 4.6236 | 1184076 | 7.6 | 0.16 | 41056 | 3.46 |
| four MIPS core + four OICs | 4.6264 | 1200320 | 10.4 | 0.22 | 57300 | 4.77 |
| eight MIPS core + one OIC | 9.2430 | 2393410 | 11 | 0.11 | 57746 | 2.41 |
| eight MIPS core + two OICs | 9.2444 | 2401532 | 12.4 | 0.13 | 65868 | 2.74 |
| eight MIPS core + four OICs | 9.2472 | 2417776 | 15.2 | 0.16 | 82112 | 3.39 |
| eight MIPS core + six OICs | 9.25 | 2434020 | 18 | 0.19 | 98356 | 4.04 |

**Table 3 Logical elements in MCS-OIC**

| | |
|---|---|
| A. Number of logical elements in One OIC | 530 |
| B. Number of logical elements in One MIPS core | 19988 |
| C. Number of logical elements in MCS-OIC (one MIPS core with one OIC) = A + B | 20518 |

The redundancy-based core level fault mitigation techniques /approaches such as Slipstream [39], Dynamic Core Coupling (DCC) approach proposed by [40] LaFrieda et al., configurable isolation [41], Reunion is a fingerprinting technique proposed by [42] Smolens et al. have nearly 100% area overhead and obviously larger power overhead. From the Table 2, the maximum power overhead and maximum area overhead are 0.46% and 11.4% respectively. It is known from Table 2, MCS-OIC (one MIPS core + one OIC) consumes least area overhead and power overhead of 0.11% and 2.41% respectively. The redundant execution of every instruction must be minimized, but still fault tolerance should be provided. MCS –OIC protects only arithmetic instructions which constitute nearly 60% of total instructions in application programs [4].

Finally, architecture of MCS-OIV in the Figure 1.2 forms the basis for reliability model discussed in the subsequent sections. The model considers L number of conventional cores (MIPS cores) and M number of OICs (L<=M) for estimating the system reliability. The model estimates the system reliability for different configurations starting from <1 MIPS core, 1 OIC> which is tightly coupled to <L MIPS cores, M OICs> loosely coupled. The OSS is adapted to capture the reliability of the computational aspect of the core. An overview on OSS model, OSS model parameters, different configurations of MCS-OIC supported, mapping of OSS parameters to MCS-OIC are presented in the next section.

## 5. OSS and Model parameters for MCS-OIC

In OSS, the system is expected to perform one or more tasks at a time. It could be repeated but it is independent of prior task execution. The OSS model presented here is for the multi-functional components (The components that can perform multiple functions are called multifunctional components). The MIPS core and OICs are multi-functional components. The model parameters of the OSS are (a) wakeup probability (b) start-up strategy and (c) readiness of the components. These parameters are briefly explained below.

**Wakeup probability:** It is the probability of successfully turning-on the function on the component. Additional power is consumed for wakeup operation.

**Start-up strategy:** Prior to system deployment, functions are tested and activated on the components and the same function is disabled on others. The wakeup probability is one for a function which is start-up strategized on a component. This aspect is significant for low power device like OICs. A function that has undergone start-up strategy does not need wake-up and saves power.

**Readiness**: It denotes the probability of readiness of the component before the operation involving migration of instructions takes place. Initially, the readiness of components in the proposed model are set to 0.99. Failures in the subsequent invocations of a component will result in reduction of readiness probability. Previous readiness is considered for evaluating system reliability when a component is requested for functional support.

Different configurations of MCS-OIC adopting the OSS model parameters are discussed in the following section.

5.1 Four different configurations of MCS-OIC

In general, a system with L number of conventional cores, M number of OICs, and N number of functions is considered. Three different configurations factoring parametric implications of OSS stated above are possible. The first configuration uses a 1:1 mapping of conventional core to OICs. In this configuration, every conventional core is allotted one OIC for functional support. The second and third configurations, use one to many mappings. In the second configuration, one conventional core accesses more than one OICs. In the third, more than one conventional core accesses multiple OICs simultaneously. However, one OIC cannot be shared by multiple cores. These configurations are explained below with respect to an example given in the Figure 1.3.

Configuration shown in the Figure 1.3 (a) is a system comprising one conventional core and one OIC with the ability to perform three functions. These three functions F1, F2 and F3 performing correctly are necessary for the system to be stable. Assuming one among the three functions is start-up strategized. Remaining two functions can be invoked on OIC with wakeup probability by the conventional cores. Configuration shown in the Figure 1.3 (b) is a system comprising two conventional cores and four OICs. Three functions F1, F2 and F3 performing correctly are necessary for the system to be stable. The four Functions (F1, F2), F3, F1, and F2 are configured in four OICs. When the function F1 fails in one among two conventional cores, OIC with high readiness and F1 startup strategized is selected to perform the task. Non-availability of such OICs leads to either another OIC with F1 startup strategized or OIC with wakeup probability selected to perform the function F1. Factoring above implications increases system reliability and the same is adopted by conventional cores to invoke two functions F3 and F2 in the configuration shown in the Figure 1.3 (b).

Configuration in the Figure 1.3 (c) is a system comprising two conventional cores and four OICs. Three functions F1, F2 and F3 performing correctly are necessary for the system to be stable. Four functions (F1, F2), F3, F2 and F1 are configured in four OICs. F2 and F1 fail simultaneously in two different conventional cores. OICs are selected to provide support to the two conventional cores. OICs with F1 and F2 startup strategized are given preference, but if they are not available, OICs with wake-probability for F1 and F2 are invoked. Configuration in the Figure 1.3 (d) is a system comprising two conventional cores and one OIC. Simultaneous access to OIC by two conventional cores is not considered in our modelling.

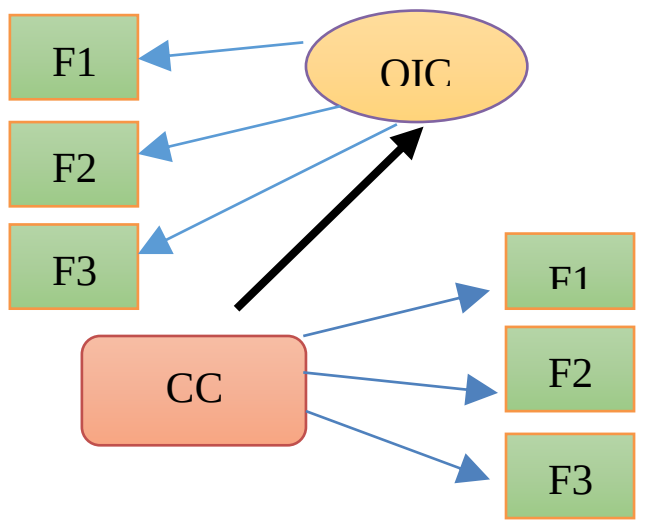

(a)

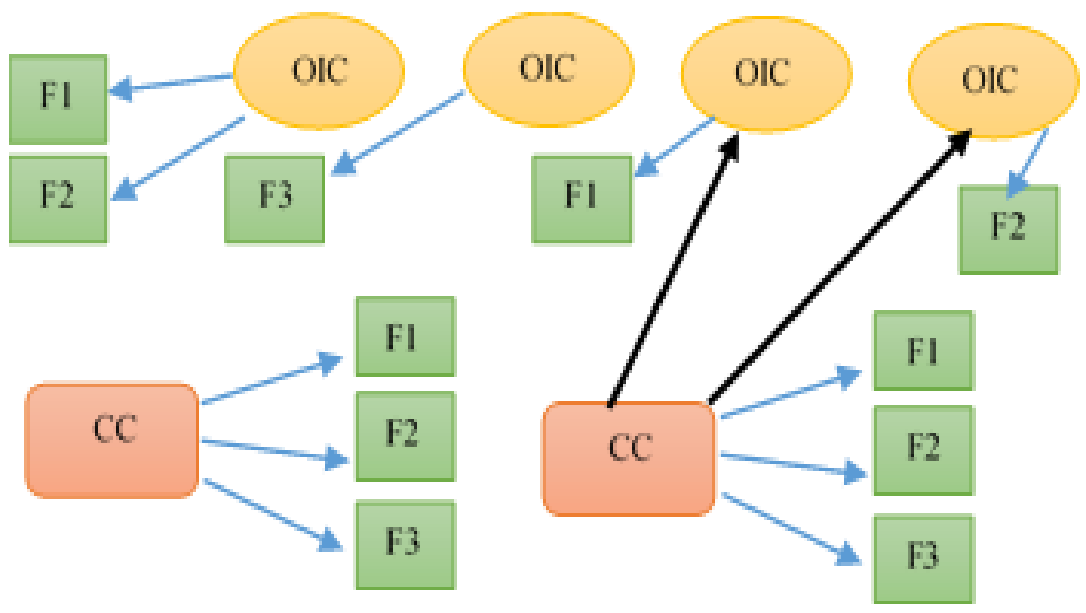

(b)

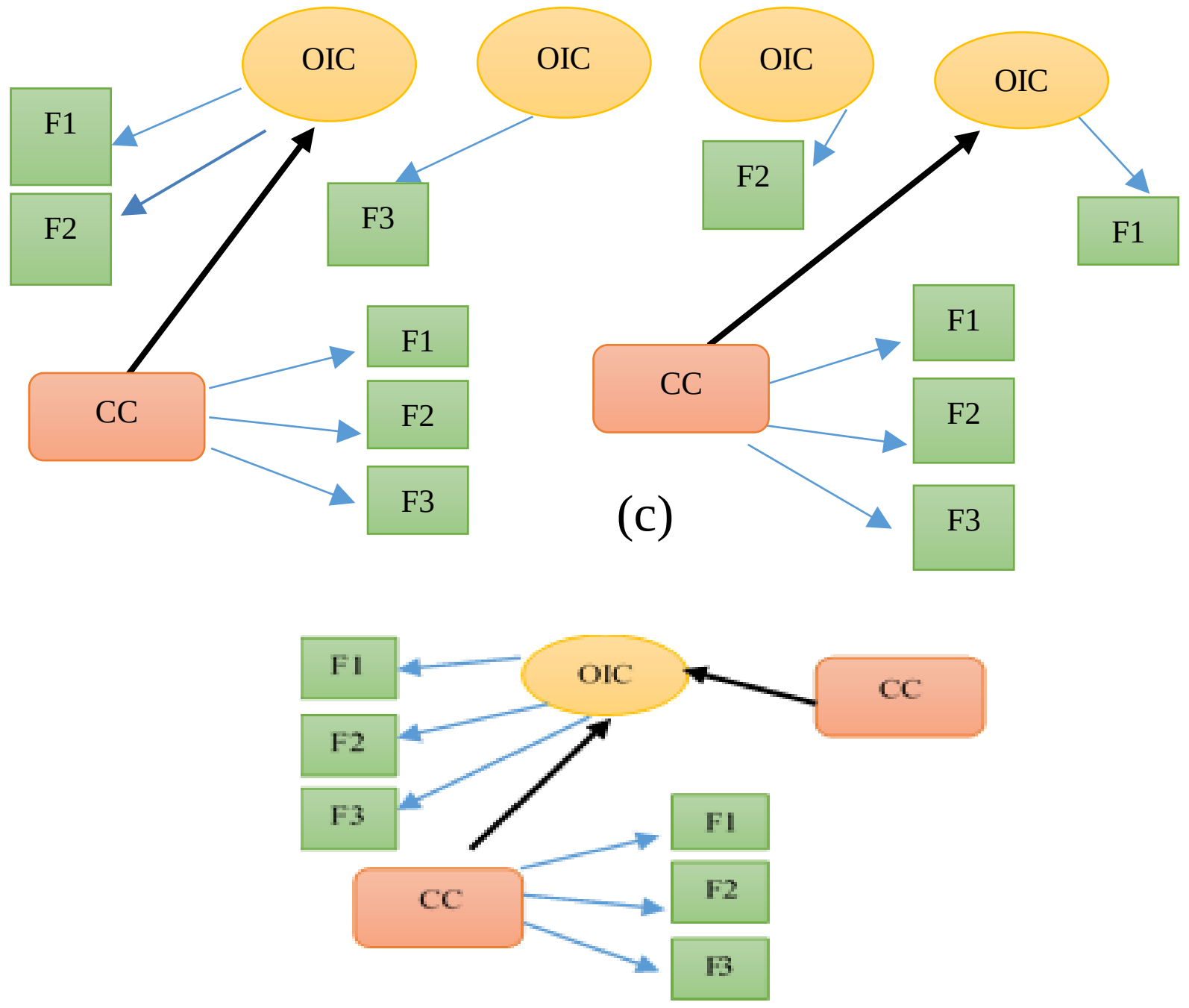

(d)

**Figure 1.3** **(a) system with one OIC supporting one Conventional core (b) Conventional core supported by two OICs, (c) two conventional cores simultaneously access two OICs (d) Simultaneous access to OIC**

5.2 OSS Model Parameters for MCS-OIC

Consider a system consisting L number of active conventional cores and M number of candidates OICs. The system is said to be working correctly if all N number of functions are performing correctly with or even without OICs. Before, the system is put to use, functions on OICs are tested and activated, at least one on each OIC. This activity is known as start-up strategy and is denoted by variable $x_{ij}$. If $x_{ij}$ is set to one, the function is ready to perform the operation. $x_{ij}$ set to zero denotes that function j is not start-up strategized.

In due course of time, conventional core seeks functional support from OICs. Availability of the function j on OIC i (denoted by $a_{ij}$) are checked before the OICs are selected for this task. $w$ number of candidates OICs are selected to support the conventional core and $k$ among them could fail to respond. $\Psi_0$ denotes the set of performing OICs. $\Psi_{k,b}$ is the index of the combinations $\binom{w}{k}$. Among the selected OICs that have component level readiness $rd_i$ and functional level readiness $r_{ij}$ are assigned to perform the task. Sometimes, OICs with high readiness $rd_i$ could fail due to lack of readiness at function level j ($r_{ij}$). The $r_{ij}$ is the probability of successfully executing the function j on the OIC or component i. OICs with similar functions with component level readiness are turned-on with wakeup probability $p_{ij}$. The reliability of the entire operation: migration of instruction and execution of instructions on OICs ($r_{ij}$) is denoted by $r(t)$. The reliability of the system is product of readiness of OICs and reliability of functions performed on OICs. The parameters are summarized in the order discussed above, in Table 4.

**Table 4 Models parameters and notations**

| Notation | Description |
|---|---|
| L | Number of independent active conventional cores in the system. |
| M | Set of candidate OICs or components $\{1,2,3 \ldots m\}$,L ≤ M |
| N | Set of functions $\{1,2,3 \ldots n\}$ |
| $x_{ij}$ | Is 1, if i[th] OIC is assigned to perform function j in start-up strategy; is 0, otherwise. |
| A | Matrix containing decision variables. $i \in M, j \in N$. |
| $a_{ij}$ | Is 1, if function j is available on i[th] OIC; is 0, otherwise. |
| $w$ | Number of candidates OICs for constructing the system, $w \leq m$. |
| $\Psi_0$ | Set of all selected OICs $\mid \Psi_0 \mid = w$ |
| $\Psi_{k,b}$ | b[th] nonempty subset of $\Psi_0$ in which $k$ OICs are not ready.$0 \leq k \leq w - 1, 1 \leq b \leq \binom{w}{k}$ and $\Psi_{0,1} = \Psi_0$. |
| $rd_i$ | Readiness level of the OIC i. |
| $p_{ij}$ | Wakeup probability: Probability of turning on function j on i[th] OIC , $i \in M, j \in N$ |
| *r(t)* | Reliability of the active core requesting for execution of the faulty Instruction. It includes $r_{ij}$ the reliability of function j on i[th] OIC. |
| $r_{ij}$ | Reliability of function level readiness j on i[th] OIC. It is an integral part of *r(t)*. |
| $S^R$ | System reliability |
| $C_{ij}$ | Cost $C_{ij}$ is the execution time in clock cycles for the function j on i[th] OIC. |
| *C* | Maximum waiting time for conventional core seeking computation from a set of OICs. |
| *U* | $<u_1, u_2, \ldots, u_N>$, $u_i$ represents the number of copies of function j in the system. It is summation of each column in A. |
| $X_{mxn}$ | Matrix containing decision variables. $i \in M, j \in N$. |
| $I_j$ | Subset of candidate OICs that are possible to perform function j. |
| sgn (z) | Is 1 if z > 0 ; is 0 otherwise. |

In the next section, theoretical reliability model for the system reliability for MCS-OIC is evaluated with simulation results and four special cases for MCS-OIC are presented.

## 6. System reliability for MCS-OIC

### 6.1 Reliability Modelling Assumptions

(a)All the OICs are independent at the component level (or its core level) and likewise, for each OIC, $rd_i$ and $p_{ij}$ are independent.

(b)The system is said to be correctly functioning, if and only if all n functions are working.

(c)All the functions of the OIC stop when the OIC is not ready.

(d)All OICs are multifunctional.

(e)All OICs are warm standby redundancy cores. OIC or component level readiness is not instantaneous.

6.2. Derivation for System Reliability

Considering the start-up strategy $X_{m\,x\,n} = [x_{ij}]$, and the availability of function j on i^th OIC, $A_{m\,x\,n} = [a_{ij}]$, the system reliability under two conditions, namely k = 0 and k > 0 are given in Equations (14) and (15) respectively. The variable *k* denotes number of failed OICs.

$$S^R = \sum_{k=0}^{w-1} \sum_{b=1}^{\binom{w}{k}} \{ [\prod_{u \in \Psi_{k,b}} rd_u] . [\prod_{v \in \Psi_0 / \Psi_{k,b}} (1 - rd_v)] . \prod_{j=1}^{n} [1 - \prod_{i \in \Psi_{k,b}} (1 - r(t) \ . E_{ij})^{a_{ij}} ] \}, \text{ when } k > 0. \tag{14}$$

$$S^R = [\prod_{u \in \Psi_0} rd_u] . \prod_{j \in N} [1 - \prod_{i \in \Psi_0} (1 - r(t) \ E_{ij})^{a_{ij}}], \text{ when } k = 0. \tag{15}$$

The derivation is presented below.

It is considered that the system is functioning when all functions are being executed using conventional cores. When few functions are not working properly in the conventional cores, OICs are selected to perform the function reliably. The term "components" or OICs are used interchangeably. Probability of all functions being performed for a given selected $\Psi_{k,b}$ OICs is given in Equation (16).

$P$ {all functions are working | $\Psi_{k,b}$ }

$= \prod_{j=1}^{n} (1 - \text{P} \{ \, all\ selected\ OICs\ fail\ to\ perform\ function\ j \ | \Psi_{k,b} \, \})$

$$= \prod_{j=1}^{n} [1 - \prod_{i \in \Psi_{k,b}} \text{Pr} \{ \, ith\ OIC\ fails\ to\ perform\ function\ j) \, \}] \tag{16}$$

In order to estimate the failure of i^th OIC failing to perform function j, $E_{ij}$ a variable is introduced which considers startup-strategy and wakeup probability and its computation is given in Equation (17).

$$E_{ij} = \begin{cases} p_{ij} \ if \ x_{ij} = 0 \\ 1 \ if \ x_{ij} = 1 \end{cases} \tag{17}$$

Now the $E_{ij}$ can be expressed using $x_{ij}$ as given in Equation (18).

$$E_{ij} = (1 - p_{ij}) x_{ij} + p_{ij} \tag{18}$$

Now the probability of i^th OIC failing to perform function j using $E_{ij}$ is expressed in Equation (19).

$$P \ ( i^{th} \text{ OIC } fails\ to\ perform\ function\ j) = (1 - r(t) E_{ij})^{a_{ij}} \tag{19}$$

**Present version is version 1.0. Manuscript is under Review. Modified or extended version will appear in the Journal.**

Considering OICs which are good ones among the selected denoted by $\Psi_{k,b}$, the probability of all functions working with given $\Psi_{k,b}$ is given in the Equation (20).

$$P \ \ (all\ functions\ are\ performing|\Psi_{k,b}\ ) = \prod_{j=1}^{n}\left[1 - \prod_{i\in\Psi_{k,b}} \ (1 - r(t)E_{ij}\ )^{a_{ij}}\right] \qquad (20)$$

The Equation (20) is arrived at by substituting Equation (19) in Equation (16). In the Equation (20), *r(t)* includes $r_{ij}$ that denotes reliability of function j on component i. Now the readiness of OICs in the system are discussed below. *w* OICs are selected and k among them failed. *k* can take the values *k = 0… w-1*. $\binom{w}{k}$ or $wC_k$ possible options are available. Let us consider b to be the ordinal number of these options. (b = 1…$\binom{w}{k}$). Let $\Psi_{k,b}$ , denote bth nonempty subset of $\Psi_0$ in which *k* OIC*s* are not ready. The range of values for k and b are: $0 \le k \le w-1,\ 1 \le b \le \binom{w}{k}$. The readiness for the combination available is given in the Equation (21).

$$Readiness\ Probability = \left[\prod_{u\in\Psi_{k,b}} rd_u\right].\left[\prod_{v\ \in\Psi_0/\Psi_{k,b}}(1 - rd_v)\right] \qquad (21)$$

$rd_u$ and $rd_v$ are the readiness of the OICs. Hence the system reliability is given in the Equation (22).

$$S^R = \sum_{k=0}^{w-1}\sum_{b=1}^{\binom{w}{k}}\{[\prod_{u\in\Psi_{k,b}} rd_u].[\prod_{v\ \in\Psi_0/\Psi_{k,b}} 1 - rd_v]\}$$
$$.\{\text{probability of all functions working with given } \Psi_{k,b}\ \} \qquad (22)$$

Substitution of Equation (20) in Equation (22) results in Equation (14) as given above.

Let us consider two examples to illustrate that the system reliability in Equation (15) is enhanced by augmenting the start-up strategized functions and function or component level redundancy. Example 1: System with three OICs and three functions in each OIC, Example 2: System with five OICs and five functions in each OIC. For both examples, Readiness $rd_i$ = 0.99998 and *r(t)* = 0.65 is assumed. The enhancement in the system reliability is observed with increase in the number of functions start-up strategized as shown in the Figure 1.4. Increase in number of OICs, number of functions in OIC, and start-up strategized functions enhance the system reliability as observed in Figure 1.5. The maximum system reliability attained is 0.9999012 and 0.99998 for three OICS with three functions and five OICs with five functions respectively. The graph is drawn with the x-axis consisting of number of functions start-up strategized and reliability on the y-axis.

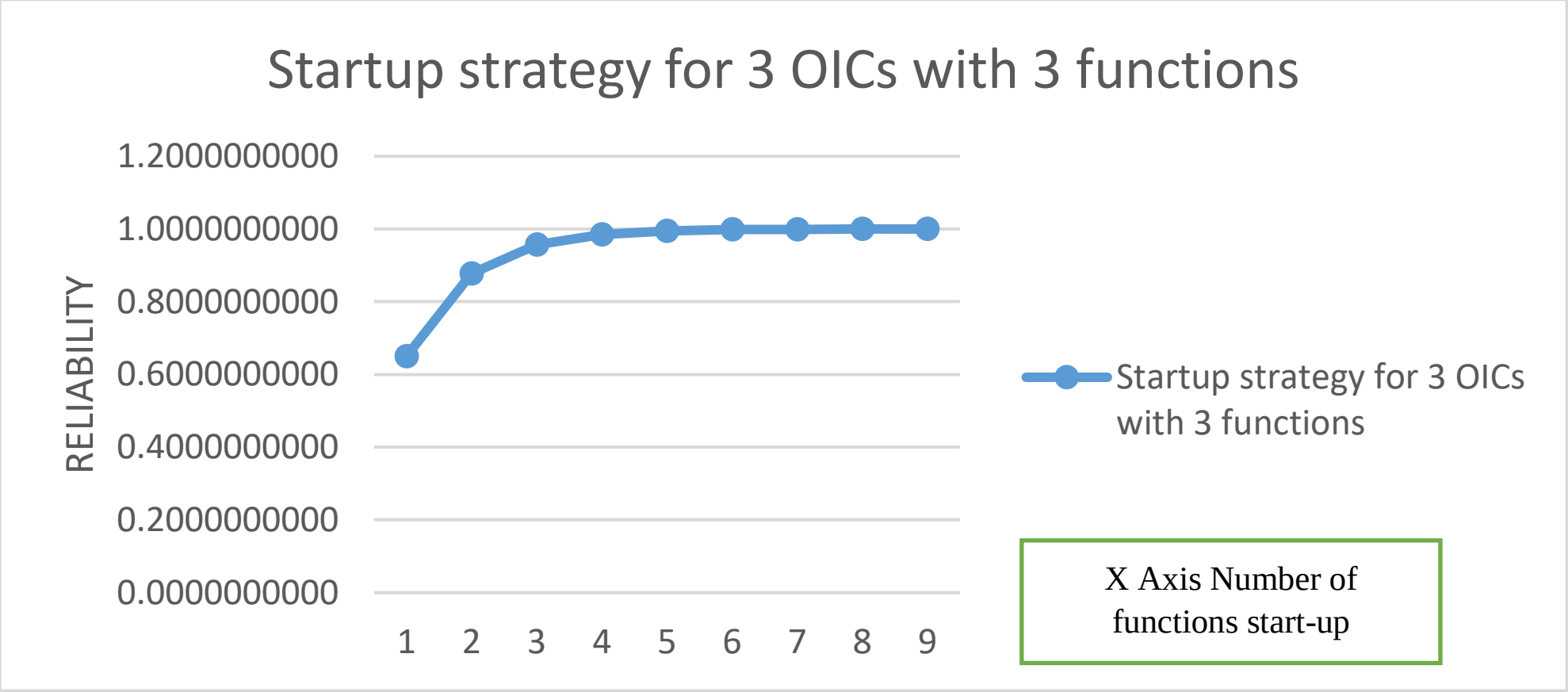

**Figure 1.4 Reliability vs Number of start-up strategized functions for 3 OICs, each with three functions.**

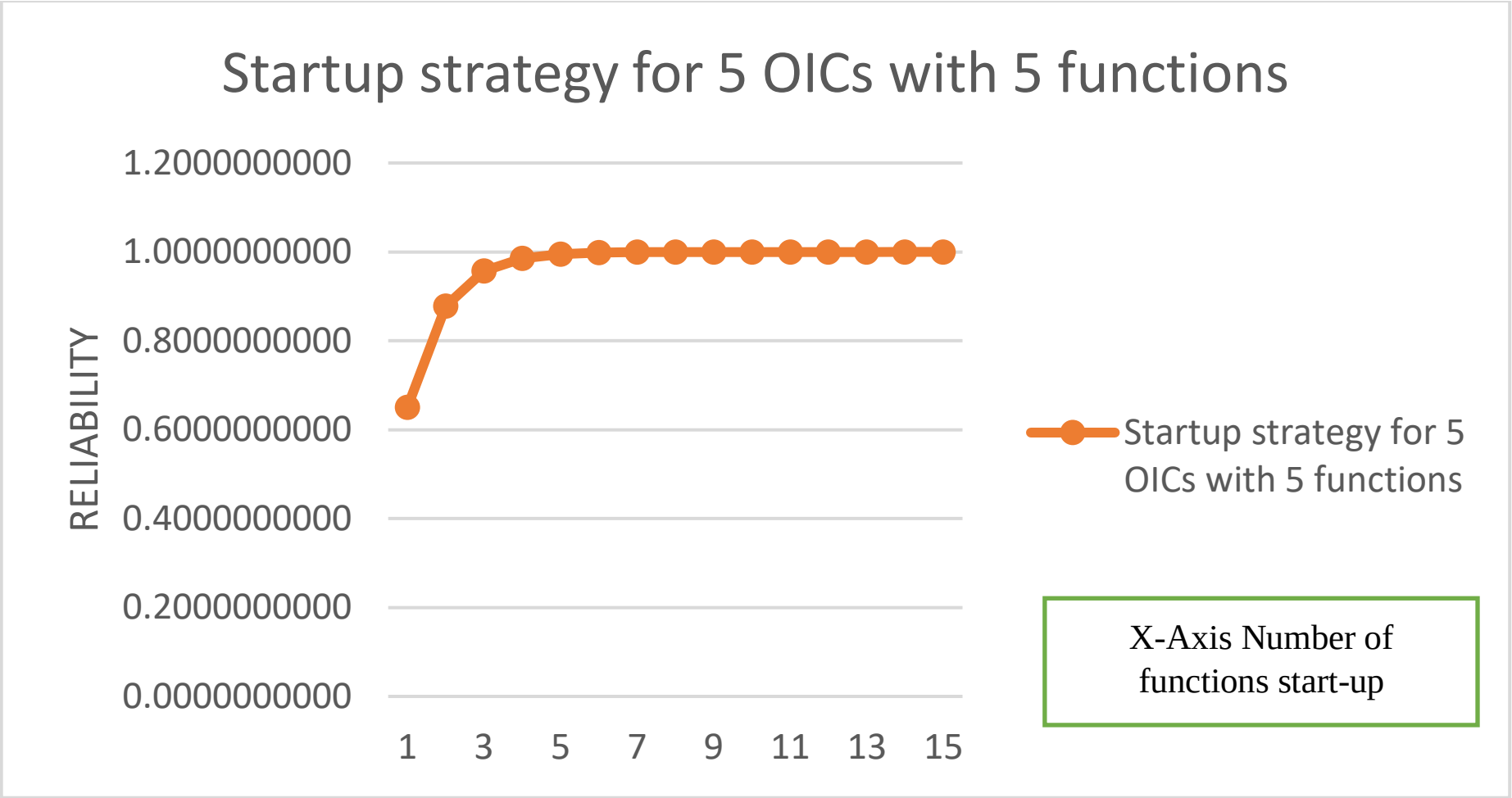

**Figure 1.5 Reliability vs Number of start-up strategized functions for 5 OICs, each with five functions.**

There are four special cases for the Equation (15), and they are (a) Readiness of the OICs are identical (b) System with number of functions ($N = M$) is equal to number of OICs with $a_{ii} = p_{ii} = 1$(c) All functions are available on all OICs with $p_{ij}$ equal to one and (d) System with identical components ($r_{ij} = R_j$ and $p_{ij} = P_j$, $j = 1, 2, 3, \ldots, n$).

Four special cases are examined below.

Special case 1:

If the Readiness of all the OICs are same, $rd_i = rd$, $i = 1, 2, 3 \ldots m$, then the reliability of the system when $k > 0$ is given in the Equation (23).

$$S^R = \sum_{k=0}^{w-1}\sum_{b=1}^{\binom{w}{k}}\{[rd^{w-k}].[(1-rd)^k].\prod_{j=1}^{n}[1-\prod_{i\in\Psi_{k,b}}(1-r(t).E_{ij})^{a_{ij}}]\} \quad (23)$$

In Equation (23), reliability of the system is estimated by assuming that readiness for all good OICs is the same. Thus, the system reliability is product of readiness of all good OICs and reliability of the active conventional core $r(t)$ which includes $r_{ij}$ (reliability of the function j on OIC i).

When $k = 0$, the system reliability is given in the Equation (24).

$$S^R = [rd^w].\prod_{j=1}^{n}[1-\prod_{i\in\Psi_{k,b}}(1-r(t).E_{ij})^{a_{ij}}] \quad (24)$$

Equation (24) computes system reliability assuming all the OICs are working and have same readiness. Considering $k = 0$ and $rd_i = rd$, system reliability in Equation (24), is a product of readiness of all OICs and reliability of the active conventional core $r(t)$, which includes $r_{ij}$ (reliability of the function j on OIC i).

Special case 2:

Under the assumption, that the system is functioning such that all functions are being executed by at least one component or core, consider a special case where number of functions is equivalent to number of components with $(m = n)$. The function j is assigned to i[th] OIC and the matrix X will be an identity matrix. i.e., $(m = n)$, $\forall\, i, j \in N(i \neq j), a_{ii} = p_{ii} = 1, a_{ij} = p_{ij} = 0$. The start-up strategy matrix X is an identity matrix and the reliability of system is derived as follows.

Derivation for Special case 2:

It is clear from the above statement that, X matrix is an identity matrix. It is known from Equation 5.5 that $E_{ij} = (1 - p_{ij})x_{ij} + p_{ij}$. Obviously, the following Equation (25) holds.

$$p_{nxn} = E_{nxn} = X_{nxn} = \begin{pmatrix} 1 & & \\ & \ddots & \\ & & 1 \end{pmatrix}, \Psi_0 = N, w = n \quad (25)$$

Since $a_{ij} = 0$ and $\forall k \neq 0, \forall i \in \Psi_{k,b}, i \neq j$, the reliability of functions represented as non-diagonal elements in the X matrix are zero and substitution of Equation (25) in Equation (14) results in Equation (26).

$$\prod_{j \in N}[1 - \prod_{i \in \Psi_{k,b}} (1 - r(t)E_{ij})^{a_{ij}}] = 0 \quad (26)$$

In Equation (26), reliability of function j on component i, where i ≠ j, is zero. For every i, j, j ∈ N $(i \neq j)$, $\Psi_0 = N, a_{ii} = 1, a_{ij} = 0,$ we have $(1 - r(t)E_{ij})^{a_{ij}} = 1$.

$$For\ i = j, we\ have\ (1 - r(t)E_{ij})^{a_{ij}} = (1 - r(t).1)^1.$$

Therefore, Equation (26) is derived as follows:

$$\prod_{j \in N}[1 - \prod_{i \in \Psi_0}(1 - r(t)q_{ij})^{a_{ij}}] = \prod_{j \in N}[1 - (1 - r(t).1\ )^1]] = \prod_{j \in N} r(t) \quad (27)$$

Substituting Equation (27) into Equation (15) results in Equation (28).

$$S^R = \prod_{j \in N} rd_j r(t) \quad (28)$$

Under these conditions, system form a series system.

Special case 3:

Considering $m = n, \forall i, j \in N, a_{ij} = p_{ij} = 1$, the reliability of the system is derived as follows:

Derivation for special case 3 :

$$p_{nxn} = E_{nxn} = \begin{pmatrix} 1 & \cdots & 1 \\ \vdots & \ddots & \vdots \\ 1 & \cdots & 1 \end{pmatrix}, \Psi_0 = N, w = n \quad (29)$$

In Equation (29), considering the assumption that number of functions = number of OICs and all functions are available on all OICs with wakeup probability $p_{ij} = 1$, $E_{ij}$ would be a matrix of ones.

Substituting Equation (29) into Equation (14) results in:

$$S^R = \sum_{k=0}^{w-1} \sum_{b=1}^{\binom{w}{k}} \{[\prod_{u \in \Psi_{k,b}} rd_u].[\prod_{v \in \Psi_0/\Psi_{k,b}} (1 - rd_v)].1\} \quad (30)$$

In the Equation (30), the system reliability is estimated when all functions are available on all OICs. Hence, the Equation (30) reduces to product of readiness of all working OICs, because reliability of all functions j on all OICs i is one.

It is known that $P$\{all components failed \} + $P$ \{ at least one component survives \}= 1.

$$\prod_{v \in N}(1 - rd_v) + \sum_{k=0}^{w-1} \sum_{b=1}^{\binom{w}{h}} \{[\prod_{u \in \Psi_{k,b}} rd_u].[\prod_{v \in \Psi_0/\Psi_{k,b}} (1 - rd_v)]\} = 1$$

$$\prod_{v \in N}(1 - rd_v) + S^R = 1$$

$$S^R = 1 - \prod_{v \in N}(1 - rd_v) \quad (31)$$

Under these conditions, the system resembles a parallel system.

Special case 4:

For the system consisting of identical OICs i.e.,$\forall i, p_{ij} = P_j, and\ X_j = \{ i | x_{ij} = 1\}$, the reliability of system is given in the Equation (32) for case *k > 0* and for case *k = 0*, the reliability of system is given in the Equation (33) .

$$S^R = \sum_{k=0}^{w-1} \sum_{b=1}^{\binom{w}{k}} \left\{[\prod_{u \in \Psi_{k,b}} rd_u].[\prod_{v \in \Psi_0/\Psi_{k,b}} (1 - rd_v)].\prod_{j=1}^{n} \left[1 - (\prod_{i \in \Psi_{k,b} \setminus X_j} \left(1 - r(t)P_j\right)^{a_{ij}} . \prod_{\in \Psi_{k,b} \cap X_j} (1 - r(t))^{a_{ij}}\right]\right\} \text{ when k >0} \quad (32)$$

$$S^R = [\prod_{u \in \Psi_0} rd_u] . \prod_{j=1}^{n} \left[1 - (\prod_{i \in \Psi_{k,b} \setminus X_j} \left(1 - r(t)\ P_j\right)^{a_{ij}} . \prod_{\in \Psi_{k,b} \cap X_j} (1 - r(t))^{a_{ij}}\right] \quad \text{when } k = 0 \quad (33)$$

According to this premise, $S^R$ can be written as follows:

$$S^R = \sum_{k=0}^{w-1} \sum_{b=1}^{\binom{w}{k}} \{[\prod_{u \in \Psi_{k,b}} rd_u].[\prod_{v \in \Psi_0/\Psi_{k,b}} 1 - rd_v].\prod_{j=1}^{n}[1 - \prod_{i \in \Psi_{k,b}} (1 - r(t).E_{ij})^{a_{ij}}]\} \quad (34)$$

where $E_{ij} = p_j\ if\ x_{ij} = 0,\ i.e., i \notin X_j,\ otherwise\ 1.$

$\prod_{i \in \Psi_{h,b}} (1 - r(t).E_{ij})^{a_{ij}}$ can be split as

$$(\prod_{i \in \Psi_{k,b} \setminus X_j} \left(1 - r(t)P_j\right)^{a_{ij}} . \prod_{\in \Psi_{k,b} \cap X_j} (1 - r(t))^{a_{ij}}) \quad (35)$$

Under these conditions system reliability is product of reliability of OICs with and without wakeup probability.

To model failure rate of active conventional core seeking functional support from OICs, Erlang distribution is used. In Erlang distribution [43], β (shape parameter) is an integer and closed form expression for reliability function is given in the Equation (36). It models one active component and β -1 identical components.

$$R(t) = \sum_{kl=0}^{\beta-1} \frac{(\lambda t)^{kl}\, e^{-\lambda t}}{kl!} \quad (36)$$

Substituting Equation (36) for *r(t)* in Equations (14) and (15) , system reliability when L = 1 is given in Equation (37) and Equation (38).

$$S^R = \int_0^{\infty} \sum_{k=0}^{w-1} \sum_{b=1}^{\binom{w}{k}} \left\{ \left[\prod_{u\in\Psi_{k,b}} rd_u\right]. \left[\prod_{v\in\Psi_0/\Psi_{k,b}} 1 - rd_v\right]. \prod_{j=1}^{n} \left\{ (1 - \prod_{i\in\Psi_{k,b}} (1 - \sum_{\beta=0}^{|\Psi_{k,b}|} \frac{(\lambda t)^{\beta}\, e^{-\lambda t}}{\beta!} . E_{ij})^{a_{ij}} \right\} \right\} dt \quad \text{when } k > 0. \quad (37)$$

In Equation (37), the system reliability when L = 1 is estimated for w selected OICs with k failures.

$$S^R = \int_0^{\infty} \left[\prod_{u\in\Psi_0} rd_u\right]. \prod_{j\in N} \left[1 - \prod_{i\in\Psi_0} \left(1 - \sum_{\beta=0}^{|\Psi_0|} \frac{(\lambda t)^{\beta}\, e^{-\lambda t}}{\beta!} \; E_{ij}\right)^{a_{ij}}\right] dt \quad \text{when } k = 0. \quad (38)$$

In Equation (38), system reliability is estimated when there are no failures in the selected OICs for L = 1 case. Reliability function in the Erlang distribution for the case L > 1, is given in Equation (39).

$$R(t) = \sum_{kl_i=0}^{\beta-1} e^{-\sum_{i=1}^{L} \lambda_i t} . \prod_{i=1}^{L} \frac{(\lambda_i t)^{kl_i}}{kl_i!} \; , i = 1, \dots, L \quad (39)$$

Thus, the system reliability with k > 0 and k = 0 for L number of independent active conventional cores are given in the Equations (40) and (41) respectively.

$$S^R = \sum_{k=0}^{w-1} \sum_{b=1}^{\binom{w}{k}} \left\{ \left[\prod_{u\in\Psi_{k,b}} rd_u\right]. \left[\prod_{v\in\Psi_0/\Psi_{k,b}} 1 - rd_v\right]. \prod_{j=1}^{n} \left[1 - \prod_{i\in\Psi_{k,b}} (1 - \sum_{kl_{ii}=0}^{|\Psi_{k,b}|} e^{-\sum_{ii=1}^{L} \lambda_{ii} t} . \prod_{ii=1}^{L} \frac{(\lambda_{ii} t)^{kl_{ii}}}{kl_{ii}!} \; . E_{ij})^{a_{ij}} \right] \right\}, ii = 1, \dots, L \quad \text{when } k > 0 \quad (40)$$

$$S^R = \left[\prod_{u\in\Psi_0} rd_u\right]. \prod_{j\in N} \left[1 - \prod_{i\in\Psi_0} (1 - \sum_{kl_{ii}=0}^{|\Psi_0|} e^{-\sum_{ii=1}^{L} \lambda_i t} . \prod_{i=1}^{L} \frac{(\lambda_{ii} t)^{kl_{ii}}}{kl_{ii}!} \; E_{ij})^{a_{ij}}\right], ii = 1, \dots, L \quad \text{when } k = 0 \quad (41)$$

The special case for Equation (41) is the system consisting of only one conventional core and one OIC to perform all j functions. Its Reliability is given in the Equation (42).

$$S^R = rd_{i=1} . \prod_{j=1}^{n} [1 - (1 - r(t)\, E_{1j})^{a_{ij}}] \quad (42)$$

Now, we examine a system reliability for a system which has more than one conventional core and one OIC, and are presented below.

The system reliability for two independent conventional cores (L=2) with (M=2, 3, 4) OICs, and for three independent conventional cores (L=3) with (M=3, 4, 5) OICs are plotted in the Figure 1.6 and 1.7, using system equations stated in the Equation (40). The λ parameter in reliability function mentioned in the Equation

(39) is a function of the number of logical elements in the circuit. Number of logical elements per OIC, per MIPS core, and per MCS-OIC (with one OIC and one MIPS core) are determined using FPGA synthesis and are tabulated in Table 3. Larger the number of logical elements in micro-architectural components, higher is the reliability of the same. Obviously, increase in number of OICs will result in enhancement in the system reliability as observed in Figures 1.6 and 1.7. **That is, enhancement in the reliability of the multi-core system using low power OICs is endorsed through the above observations.**

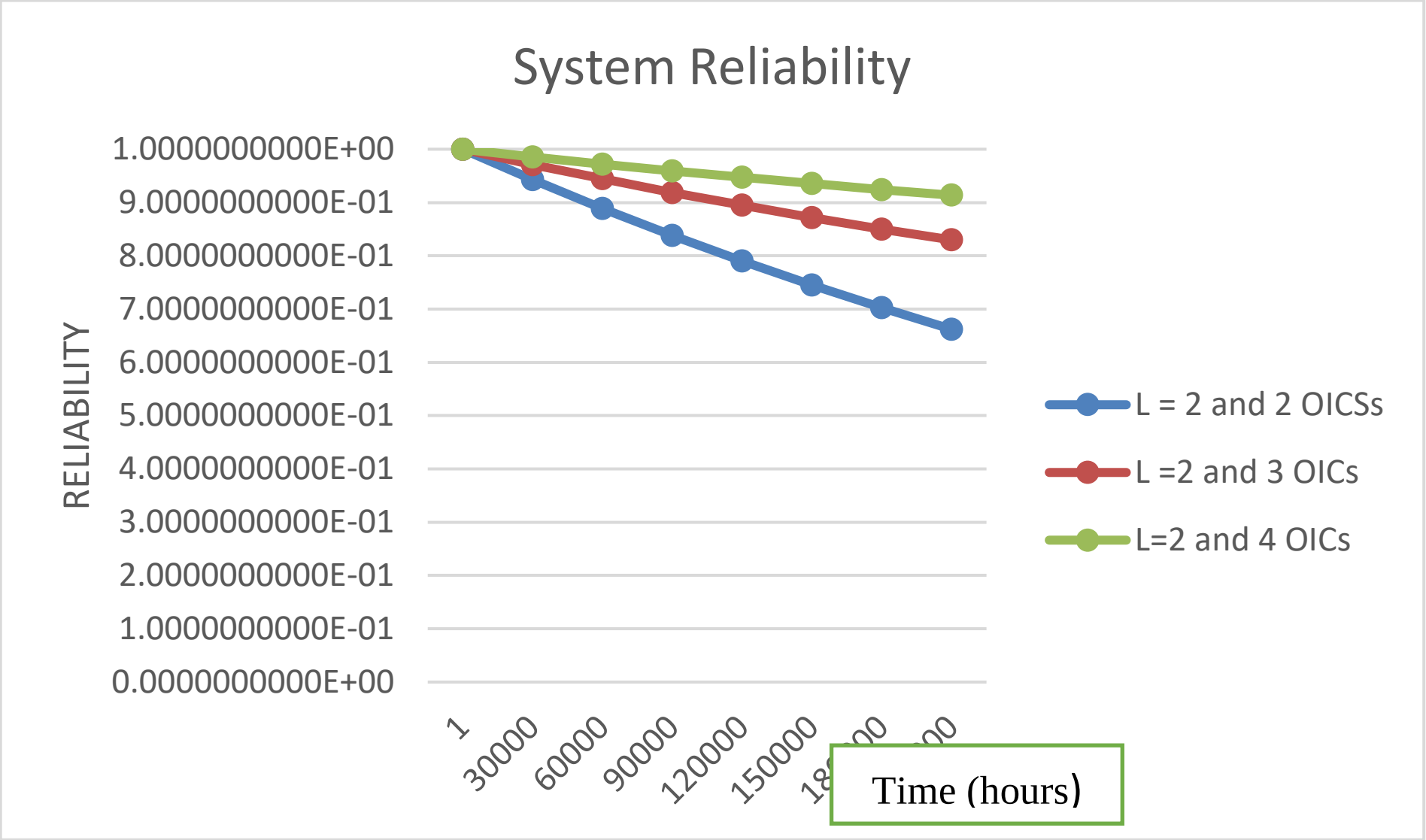

**Figure 1.6 System reliability vs Time (hours) for multi-core system with 2 independent conventional cores and 2, 3, 4 OICs.**

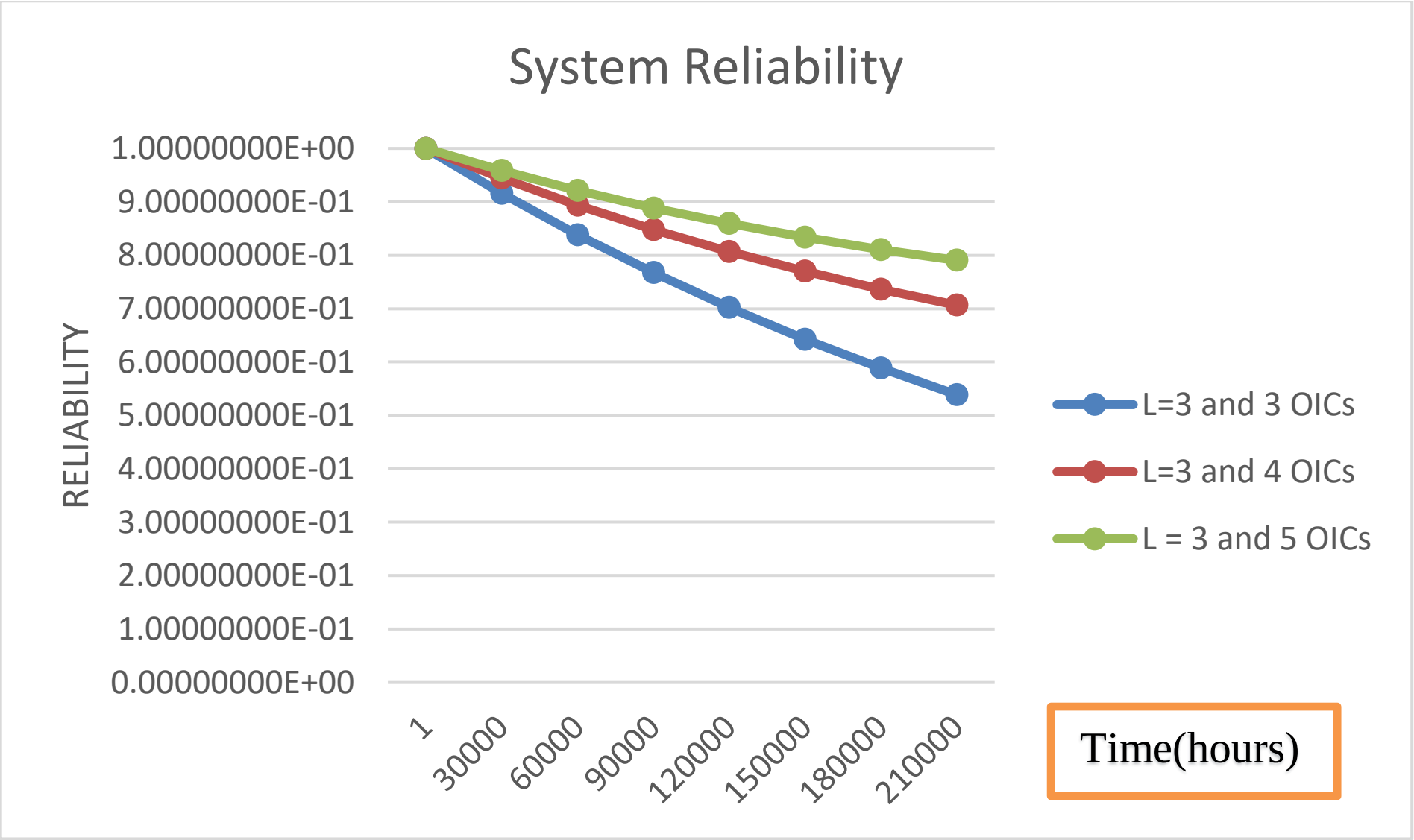

**Figure 1.7 System reliability vs Time (hours) for multi-core system with 3 independent conventional cores and 3, 4, 5 OICs.**

# 7. Interval RAP and Its Inherent Characteristics

As examined in the section V, in a MCS-OIC, there are three configurations for mapping of functions in OICs to conventional cores. In the first configuration, which is a 1:1 mapping, functions in the OIC are directly mapped to the conventional core. In the second and third configurations, which involve one to many mapping, choice of OICs and their functions that are required by conventional cores, becomes an important consideration that determines the reliability of the system. This choice of OICs and their functions is viewed as a redundancy allocation problem. Formulation for RAP is presented in this section with its inherent properties.

As examined in the previous section, an increase in the system reliability in the OSS based model is observed, when number of functions start-up strategized ($X$) in OICs and function level or component level redundancy ($A$) increases. However, there is a cost associated in terms of performance overheads in executing the functions on the OIC. Hence, the system reliability has to be optimized by determining values for X and A with the overhead cost C as the constraint. Further, another factor to be considered is that the reliability of migrating instructions from conventional core to OICs and executing the function on OICs, $r(t)$ , is imprecise. This impreciseness could be due to manufacturing complexity, or complexity in design and the environment where it is deployed. Interval analysis is used to address impreciseness. Hence Interval valued reliabilities are assigned to $r(t)$. Thus, the RAP is referred to as Interval RAP.

## 7.1. Formulation for Interval RAP

Equation (15) is restated below in Equation (43) , is the objective function to maximize the system reliability with decision variables X and A, with no failure in the selected OICs.

Maximize

$$S^R(X,A) = \left[\prod_{i\in\Psi_0} rd_i\right] . \prod_{j\in N}\left[1 - \prod_{i\in\Psi_0}(1 - r(t)E_{i,j})^{a_{i,j}}\right] rd_i , r(t) \in [0,1], 0 < t_l < t < t_r . \text{when } L \le M \text{ and } k = 0 \quad (43)$$

Subject to the constraints:

$$E_{ij} = (1 - p_{ij})x_{ij} + p_{ij}, p_{ij} \in [0,1] \quad (44)$$

$$\sum_{i=1}^{m}\sum_{j=1}^{n} C_{ij}\, x_{ij} \le C \quad (45)$$

$$x_{ij} \in [0,1] \quad (46)$$

The decision variable X denotes the start-up strategy $x_{ij}$ and A denotes the availability $a_{ij}$. In the nonlinear constraint (Equation (45)), the cost $C_{ij}$ is the execution time in clock cycles for the function j on component i. C denotes the maximum waiting time for a conventional core seeking computation from the given set of OICs. $t \in [t_l, t_r]$ is the interval time. The event defining conventional core seeking functional support from OIC may vary within interval time and is called interval mission time. Within the interval, with mission time reaching the lower bound, the OIC reliability may reach the upper bound. With the mission time reaching the upper bound, reliability of the OIC may reach lower bound. It implies that for t = $t_L$, r ($t_L$) = $r_R$ and for t = $t_R$ , r($t_R$) = $r_L$. Now, the objective function can be restated as per Definition 1.2 and is given in Equation (47). The formulation in Equation (47) is a non-linear integer programming problem with interval coefficients.

Maximize

$$S^R(X,A) = \begin{bmatrix} \prod_{i\in\Psi_0} rd_i \cdot \prod_{j\in N}\left[1 - \prod_{i\in\Psi_0}(1 - r_R E_{i,j})^{a_{i,j}}\right], \\ \prod_{i\in\Psi_0} rd_i \cdot \prod_{j\in N}\left[1 - \prod_{i\in\Psi_0}(1 - r_L E_{i,j})^{a_{i,j}}\right] \end{bmatrix} \quad (47)$$

and the constraints remain the same.

B. Important Characteristics of the Proposed RAP System

Important characteristics integral to the RAP are discussed below.

(a) Monotonicity: $S^R(x_{ij},a_{ij})$ is monotonically increasing with $a_{ij}$ $(i = 1 \ldots m;\ j = 1 \ldots n)$.

Proof:

We know that $r_{ij} \in [0,1]$, $E_{ij} \in [0, 1]$, and $r(t) \in [0,1]$. $r_{ij}$ is an integral part of $r(t)$ and $a_{ij} \geq 0$. It is known that $(1 - r(t)E_{ij})^{aij} \in [0,1]$. $i^*$ and $j^*$ which are arbitrary values of i and j, are fixed and not varied. For two instances $a^{(2)}_{i^*j^*}$, $a^{(1)}_{i^*j^*}$, if $a^{(2)}_{i^*j^*} > a^{(1)}_{i^*j^*}$ then two conditions S1 and S2 arise which are examined below.

S1: Consider $a^{(1)}_{i^*j^*} \geq 1$ *and that* $\Psi^{(2)}_0 = \Psi^{(1)}_0$, subsets of selected components of both $\Psi^{(2)}_0$ and $\Psi^{(1)}_0$ are same. Consider $F \subset \Psi^{(1)}_0 = \Psi^{(2)}_0$. For the F which does not contain $i^*$, the system reliability determined using Equation (48) for $a_{ij} = a^{(2)}_{i^*j^*}$ and $a_{ij} = a^{(1)}_{i^*j^*}$ are identical. But, for the F that contains $i^*$ and $a^{(2)}_{i^*j^*} > a^{(1)}_{i^*j^*}$, the inequality in the Expression (49) holds. Hence, the sum of all subsets F containing $a^{(2)}_{i^*j^*}$ is greater than or equal to F containing $a^{(1)}_{i^*j^*}$.

$$S^R = \left[\prod_{u\in F} rd_u\right].\left[\prod_{v\in\Psi_0\backslash F}(1 - rd_v)\right].\prod_{j=1}^{n}\left[1 - \prod_{i\in F}(1 - r(t)E_{ij})^{a_{ij}}\right] \quad (48)$$

$$(1 - r(t)^* E_{i^*j^*})^{a^{(2)}_{i^*j^*}} \leq \left(1 - r(t)^* E_{i^*j^*}\right)^{a^{(1)}_{i^*j^*}} \quad ,where\ r_{i^*j^*} \text{ is an integral part of r(t)*} \quad (49)$$

S2: Consider $a^{(1)}_{i^*j^*} = 0$, then $a^{(2)}_{i^*j^*}$ is greater than zero, and $i^* \in \Psi^{(2)}_0$. But $i^*$ does not belong $\Psi^{(1)}_0$. For $F^{(2)} \subset \Psi^{(2)}_0$ and $F^{(1)} \subset \Psi^{(1)}_0$, two conditions S3 and S4 are analysed and are discussed below.

S3: $i^*$ does not belong to $F^{(2)}$ *and* $F^{(1)}$. The system reliability determined using $F^{(2)}$ *and* $F^{(1)}$ are same and is given in Equation (50) .

$$\left[\prod_{u\in F^{(2)}} rd_u\right].\left[\prod_{v\in\Psi_0\backslash F^{(2)}}(1 - rd_v)\right].\prod_{j=1}^{n}\left[1 - \prod_{i\in F^{(2)}}(1 - r(t)E_{ij})^{a_{ij}}\right] =$$

$$\left[\prod_{u\in F^{(1)}} rd_u\right].\left[\prod_{v\in\Psi_0\backslash F^{(1)}}(1 - rd_v)\right].\prod_{j=1}^{n}\left[1 - \prod_{i\in F^{(1)}}(1 - r(t)E_{ij})^{a_{ij}}\right] \quad (50)$$

S4: $i^*$ belongs to $F^{(2)}$ such that $F^{(2)} = F^{(1)} \cup \{i^*\} \subset \Psi_0^{(2)}$ and $F^{(1)} \subset \Psi_0^{(1)}$. System reliability determined using $F^{(2)}$ is greater than or equal to system reliability using $F^{(1)}$. Then following inequality in the Expression (51) holds.

$$[\prod_{u\in F^{(2)}} rd_u].[\prod_{v\in \Psi_0 \backslash F^{(2)}}(1- rd_v)].\prod_{j=1}^{n}[1-\prod_{i\in F^{(2)}}(1-r(t)E_{ij})^{a_{ij}}] \geq$$

$$[\prod_{u\in F^{(1)}} rd_u].[\prod_{v\in \Psi_0 \backslash F^{(1)}}(1- rd_v)].\prod_{j=1}^{n}[1-\prod_{i\in F^{(1)}}(1-r(t)E_{ij})^{a_{ij}}] \quad (51)$$

It is obvious from Equation (51), that system reliability determined using $F^{(2)}$ is greater than or equal to system reliability resulting by $F^{(1)}$. Additional component $i^*$ is included in $F^{(2)}$. **Enhancement in the system reliability is proved.**

Explanation for proof:

Assuming two instances $a_{i^*j^*}^{(2)}$, $a_{i^*j^*}^{(1)}$ and $\Psi_0^{(1)} = \Psi_0^{(2)}$ holds. In both the instances (2) and (1), number of selected components are same, then the reliability computed using Equation (48) are same. If F of instance (2) contains $i^*$, and $a_{i^*j^*}^{(2)} > a_{i^*j^*}^{(1)}$, then the inequality (49) holds. Assume $i^*$ component does not belong to both $F^{(2)}$ *and* $F^{(1)}$, then reliability computed using $F^{(2)}$ *and* $F^{(1)}$ in Equation (50) are same. Assume $F^{(2)} = F^{(1)} \cup \{i^*\} \subset \Psi_0^{(2)}$ holds, then the system reliability computed using $F^{(2)}$ is greater than or equal to system reliability computed using $F^{(1)}$ as given in the Equation (51).

6.3 Reliability of Function j:

The reliability of the function j can be evaluated for the selected components and is given in Equation (52).

$$r_{fj} = 1 - \prod_{i\in I_j}[1- rd_i\,(1-(1-r_{ij}.E_{ij})^{a_{ij}}] \quad (52)$$

$r_{ij}$ is an integral part of $r(t)$. $I_j$ represents the index set of the components that can perform function j.

6.4 Maximum System Reliability:

Assuming $A(a_{ij})$ is known for a system, in order to maximize the system reliability, best start-up strategy is initialized to one, i.e. $x_{\tilde{\imath}_j j} = 1$ for function j where

$$\tilde{\imath}_j = arg \max_{i\in\Psi_0} \left\{1-\prod_{l\in I_j}[1- rd_l(1-(1- r_{lj}.E_{lj}^{(i)})^{a_{ij}})\ ]\right\} E_{lj}^{(i)} = 1, if\ l = i; otherwise\ p_{lj} \quad (53)$$

## 8. Genetic Algorithm for Interval RAP

Genetic algorithm is proposed for the nonlinear integer programming with the objective function given in the Equation (47) as the fitness function. The stochastic based selection procedure, crossover, mutation and elitism are the computational techniques adopted in the proposed GA. The proposed GA has two phases: primary phase and the secondary phase. In the primary phase, d independent runs are performed to initialize the solutions in the initial population. The solutions from the runs with subsequent consecutive solutions effectively improving the system reliability are included in the initial population. This approach of choosing the initial population will set solution evolution in the right direction and improve the efficiency of the GA. Notably, the regular GA without elitism uses the randomly initialized population to evolve solutions.

Two problems, evaluation example-one with smaller problem space and evaluation example-two with larger problem space are evaluated using GA. For evaluation example-two or large problem, additional changes are made to the primary phase of the proposed GA to highlight the effectiveness of the proposed GA as compared to regular GA. They are discussed below.

Firstly, for maximizing the system reliability, A ($a_{ij}$) is estimated without considering into $x_{ij}$. Then the start-up strategy $x_{ij}$ is set according to Equation (53). The reliability of the function j without considering a start-up strategy can be determined by replacing $E_{ij}$ by $p_{ij}$ in Equation (52) and is given in Equation (54).

$$r_{fj} = 1 - \prod_{i\in I_j}\left[1 - rd_i\,(1 - (1 - r_{ij}.p_{ij})^{a_{ij}}\right] \qquad (54)$$

Secondly, estimating $A_{ij}$ involves two steps and they are as follows.

1. Choosing components that increase functional reliability ($r_{fj}$) is performed by enabling functions such that maximum latency is within C. 2. Choosing between two OICs (say a, b), enabling one function in 'a' and disabling the same in 'b' such that $r_{fj}$ improvement is maximal is selected. A ($a_{ij}$) is determined and set to X according to $\tilde{\imath}_j$ as in Equation (53).

Finally, the solution obtained in the primary phase is included in the initial population for the secondary phase of the proposed GA.

The secondary phase of the proposed GA is described below.

1. The GA parameters, P_SIZE: population size, P_MUTAT: the probability of mutation, P_CROSS: the probability of crossover and M_GEN: number of generations are initialized. The encoding of the chromosome for the problem is given in Figure 1.8. The chromosome comprises of decision variables X (which denotes the start-up strategy) and U (which denotes the number of functions of each type used). Number of ones in the binary encoded string represent the functions that are start-up strategized (X) as shown in Figure 1.8

| | |
|---|---|
| < ----------- X --------------------------------- > | < ---------------- U ----------- > with interval values |
| 111000 001101 111000 | 1 2 0 1 0 3 1 |

**Figure 1.8 Example for chromosome encoding**

2. Each chromosome is evaluated for fitness. The fitness function is used to select the best chromosome in the population. The value of the interval valued objective function of the chromosome is taken as the fitness function for RAP systems. Only chromosomes with fitness value greater than a threshold value are considered for selection.

3. Selection: The primary objective is to evolve better solutions and eliminate below average solutions from the current population for the next generation. In the proposed GA, individuals are stochastically selected from the current population according to their fitness value. Definitions 1.3 and 1.4 are appropriately used to determine the reliability intervals for the selection of quality chromosomes.

4. Crossover for chromosomes/individuals with probability (P_CROSS) is allowed to participate to produce offspring. The 2-point crossover operator is used.

5. New species are not suddenly created. Hence, the mutation probability P_MUTAT is chosen to create new population. The mutation creates new diversity in the population.

6. Elitism: the best optimal solution is retained in all iterations to improve the performance of GA.

7. The iterations are repeated until the maximum optimal solution with respect to reliability and cost are attained. The objective function is evaluated to determine the system reliability. The best results of the system reliability at $j^{th}$ and $(j+1)^{th}$ generation are compared and superior among them are accepted. The steps 1 to 7 are repeated until changes in the system reliability are insignificant.

## 8.1 Evaluation for Genetic Algorithm

The initialization and evaluation of model parameters of OSS used in the estimation of system reliability are examined here. The following is also applicable for evaluation of PSO. (a)Readiness parameter $rd_i$: In a multi-core system involving OICs, component level readiness is high in the initial stages of operation. Hence, $rd_i$ is set to 0.99. In the subsequent requests, previous readiness is considered for evaluating system reliability. (b)Wakeup probability $p_{ij}$: In the OICs, wakeup probability of functions that have undergone start-up strategy are set to one. The functions that have not undergone start-up strategy will have a wakeup probability between 0.1 and 0.9. Functions on OICs turned on with in a time period will have higher wakeup probability compared to the other functions which fail to meet the dead line. Thus, same function on different OICs may have different wakeup probability.(c) Parameter $r(t)$ (which includes $r_{ij}$): The interval values are assigned due to impreciseness in operation that includes migration and execution of instruction on OICs. The sample set of interval values given in the Table 5 are taken from the work by Roy et al. [44]. This interval value set has been used (a) to study system entropy [44], and (b) to study maximum load point variations and voltage stability in electric power systems [31]. The significance of this set of intervals valued numbers is that it contains all types of intervals: disjoint, partially overlapping and one interval contained in the other. Disjoint interval numbers are present in the set 1 given in Table 5. A mixture of disjoint and partial overlapping interval numbers is present in sets 2-4 given in Table 5. One interval number contained in the other are present in interval numbers set 5 in Table 5. These sets of interval numbers will oscillate the decision making between optimistic and pessimistic for maximization of system reliability. The stability of the system is put to test in a realistic sense and is examined in the sensitivity section.

Two problems are considered for evaluation – “evaluation example-one” and “evaluation example-two large” as mentioned earlier. In evaluation example one, six functions and three OICs is considered. In the evaluation example two, 10 functions and six OICs is considered.

## 8.2 Evaluation example-one

In the primary phase, ten independent runs are used to initialize the solutions in the initial population. The evaluation example-one with six functions and three OICs is evaluated by performing 20 independent runs in the secondary phase of the proposed GA for each interval number set stated in Table 5. The six functions supported by OICS are ADD, MOV, INC, DEC, SUB, and DIV. The population size is initialized to 100 and the number of generations to 200. The two-point crossover operator is used to generate new offspring. Over 80% of the population undergo two-point crossover operation. 6% of individuals are mutated. The wakeup probability and readiness parameters are tabulated in Table 6. The cost $C_{ij}$, the execution time in clock cycles for the function j on component i is shown in Table 6. In Equation (45), C is set to 50 clock cycles for the evaluation example one.

The GA parameters like selection function, crossover and mutation are set similarly for both proposed GA and Regular GA. The best results for system reliability in the form of System Reliability (Lower) and System Reliability (Upper) are tabulated in Table 7. The optimal solution given by the proposed GA and regular GA are shown in the Table 8. The proposed GA provides better system reliability (lower/upper) solution compared to regular GA for all data samples in Table 8. From Table 8, in the first column, X: 101000 000010 101101, first six bits denote that ADD and INC functions in the first OIC are start-up strategized functions; second six bits denote that SUB function in second OIC is a start-up strategized function; third six bits denote that ADD, INC, DEC, and DIV functions in third OIC are start-up strategized functions. From Table 8, in first column, U: 2 2 2 3 3 3 denotes number of copies of functions ADD, MOV, INC, DEC, SUB and DIV available in the system. For a given X and U in the Table 8, the optimal system reliability (lower/upper bound) attained by proposed GA is 0.969002/0.970178 for C=44 clock cycles. Similarly other entries in the Table 8 can be interpreted.

From Table 7, the gap between system reliability (upper/lower bound) values given by regular GA and proposed GA respectively is more visible for all the samples given in Table 5. From Table 8, difference in the near optimal solutions given by both GAs concludes that primary phase has improved the efficiency of proposed GA.

**Table 5 Interval Numbers ($r_R$, $r_L$)**

| Interval number Set No. | Type of Interval numbers | Interval numbers formatted as [lower bound, upper bound] |
|---|---|---|
| SET 1 | Disjoint intervals | [0.68,0.72], [0.73,0.75], [0.78,0.81], [0.80,0.88], [0.89,0.95] |
| SET 2 | Disjoint and Partially overlapped intervals | [0.65,0.70], [0.71,0.73], [0.80,0.88], [ 0.82,0.87], [0.90,0.92] |
| SET 3 | Disjoint and Partially overlapped intervals | [0.60,0.67], [0.72,0.78], [0.78,0.83], [0.80,0.90], [0.87,0.956] |
| SET 4 | Disjoint and Partially overlapped intervals | [0.64,0.65], [0.72,0.74], [0.80,0.88], [0.83,0.85], [0.90,0.95] |
| SET 5 | One interval contained in the other, Disjoint and Partially overlapped intervals | [0.63, 0.66], [0.64, 0.68], [0.65, 0.70], [0.65, 0.70], [0.73, 0.74], [0.75, 0.79], [0.76, 0.86],[0.75, 0.80], [0.77, 0.80], [0.75, 0.81], [0.78, 0.84], [0.80, 0.87], [0.88, 0.92], [0.89, 0.90],[0.91, 0.96] |

**Table 6 Cost, Wakeup probability and Readiness parameters for evaluation example-one**

| Cost($C_{ij}$) | Wakeup probability($p_{ij}$) | Readiness $rd_i$ |
|---|---|---|
| 4 5 4 1 1 35 | 0.98 0.9 0.9 0.96 0.87 0.87 | $rd = rd_1 = rd_2 = rd_3 = 0.99$ |
| 4 5 4 1 1 35 | 0.82 0.82 0.82 0.9 0.9 0.9 | |
| 4 5 4 1 1 35 | 0 0.9 0.9 0.9 0.9 0.9 | |

**Table 7 Best System reliability (Lower/ Upper) for the evaluation example-one with 3OICs and 6 functions using GA**

| **Proposed GA** | **Table 6.1 (set 1)** | **Table 6.1 (set 2)** | **Table 6.1 (set 3)** | **Table 6.1 (set 4)** | **Table 6.1 (set 5)** |
|---|---|---|---|---|---|
| System Reliability (Lower)(Best) | 0.969006 | 0.969328 | 0.968163 | 0.969328 | 0.969591 |
| System Reliability (Upper)(Best) | 0.970178 | 0.969802 | 0.970178 | 0.970178 | 0.970237 |
| Regular GA | Table 6.1 (set 1) | Table 6.1 (set 2) | Table 6.1 (set 3) | Table 6.1 (set 4) | Table 6.1 (set 5) |
| System Reliability (Lower)(Best) | 0.968975 | 0.968452 | 0.967877 | 0.968454 | 0.962434 |
| System Reliability (Upper)(Best) | 0.970132 | 0.969230 | 0.970132 | 0.969947 | 0.968746 |

**Table 8 Optimal solution for evaluation example-one with 3OICs and 6 functions using GA**

| **Proposed GA** | **Table 6.1 (set 1)** | **Table 6.1 (set 2)** | **Table 6.1 (set 3)** | **Table 6.1 (set 4)** | **Table 6.1 (set 5)** |
|---|---|---|---|---|---|
| **X** | 101000<br>000010<br>101101 | 110100<br>000111<br>011000 | 010000<br>000011<br>110111 | 111011<br>000110<br>001000 | 111110<br>001111<br>011110 |
| **A** U:<br>OICs | 2 2 2 3 3 3<br>3 | 2 2 2 3 3 3<br>3 | 2 2 2 3 3 3<br>3 | 2 2 2 3 3 3<br>3 | 1 2 3 3 3 1<br>3 |
| System Reliability Lower/Upper | 0.969002/ 0.970178 | 0.968437/ 0.969224 | 0.958130/ 0.966943 | 0.951516/ 0.963214 | 0.962376/ 0.968744 |
| Cost (in cycles) | 44 | 40 | 46 | 49 | 41 |
| Regular GA | Table 6.1 (set 1) | Table 6.1 (set 2) | Table 6.1 (set 3) | Table 6.1 (set 4) | Table 6.1 (set 5) |
| X | 010111<br>000001<br>111111 | 111101<br>000111<br>101111 | 111101<br>000110<br>011111 | 011011<br>000111<br>101001 | 110110<br>000010<br>111101 |
| A U:<br>OICs | 2 2 2 3 3 3<br>3 | 2 2 2 3 3 3<br>3 | 2 2 2 3 3 3<br>3 | 2 2 2 3 3 3<br>3 | 2 2 3 3 3 1<br>3 |
| System Reliability Lower/Upper | 0.968780 0.970131 | 0.960412 0.964004 | 0.953624 0.967867 | 0.968101 0.968977 | 0.958415 0.967867 |
| Cost(in cycles) | 41 | 49 | 49 | 45 | 49 |

## 8.3 Evaluation example-two

The evaluation example-two with 10 functions and six OICs is evaluated by 10 independent trials of 60 runs each with 20000 generations for data samples in Table 5. The 10 functions are designated as F1, F2, F3,…, F10. The wakeup probability, cost $C_{ij}$ (and C = 3000 cycles) and readiness parameters are tabulated in the Table 9.

The best system reliability (lower/upper) results for the evaluation example two problem is tabulated in Table 10. The optimal solution by the proposed GA for the evaluation example two are shown in the Table 11. From the Table 11, X: 1 0 0 0 0 0 1 1 1 0, 0 0 0 0 0 0 1 1 0 0, 0 0 0 0 0 0 0 0 0 1, 0 1 0 0 0 0 0 0 0 0, 0 1 0 0 1 1 0 1 1 0 , 0 0 0 0 1 0 0 0 0 1, first ten bits denote F1, F7, F8, and F9 are start-up strategized in the first OIC. Similarly remaining bit strings can be interpreted. From Table 11, U: 3 3 3 1 3 3 3 3 4 4 denotes number of copies of functions F1, F2, F3,…, F10 available in the system. For a given X and U in the Table 11, the optimal system reliability (lower/upper) attained by proposed GA is 0.989966 / 0.989997 for C=2500 clock cycles. Similarly other entries in the Table 11 can be interpreted. The solutions for system reliability by proposed GA are compared with the regular GA. From the Table 10 and Figure 1.9, the solution of the proposed GA converges faster as compared to the regular GA. The proposed GA stabilizes at 0.989998/0.990000 against 0.989966/0.989997 given by the regular GA. The faster convergence of the solution in large problem space is due to changes made in the primary phase of the proposed GA and is evident from Figure 1.9. From Table 11, the system reliability (lower/upper) given by proposed GA is 0.989966 / 0.989997, which is better than 0.989876 / 0.989991 given by regular GA.

**Table 9 Cost, Wakeup probability and Readiness parameters for evaluation example-two**

| | |
|---|---|
| Cost $C_{ij}$ | 100 100 250 250 500 500 600 600 800 800<br>100 100 0 250 500 500 600 600 800 800<br>100 100 250 0 500 500 600 600 800 800<br>100 100 250 250 500 500 600 600 800 800<br>100 100 250 250 500 500 0 600 800 800<br>100 100 250 250 500 500 0 600 800 800 |
| Wakeup probability($p_{ij}$) | 0.9 0.7 0.7 0.9 0.9 0.7 0.8 0.8 0.8 0.7<br>0.9 0.7 0 0.7 0.9 0.9 0.8 0.8 0.8 0.7<br>0.9 0.7 0.9 0 0.7 07 0.8 0.8 0.8 0.7<br>0.9 0.7 0.9 0.7 0.9 0.7 0.8 0.8 0.8 0.7<br>0.9 0.8 0.8 0.9 0.9 0.9 0 0.8 0.8 0.8<br>0.9 0.8 0.8 0.9 0.9 0.8 0 0.8 0.8 0.8 |
| Readiness $rd_i$ | $rd_1$=$rd_2$ = $rd_3$=$rd_4$=$rd_5$=$rd_6$=0.99. |

**Table 10 Best System reliability (Lower/ Upper) for the evaluation example-two with 6 OICs and 10 functions using GA**

| Proposed GA | System Reliability (Best) (Lower) | System Reliability (Best) (Upper) | Regular GA | System Reliability (Best) (Lower) | System Reliability (Best) (Upper) |
|---|---|---|---|---|---|
| Table 6.1 (set 1) | 0.989998 | 0.990000 | Table 6.1 (set 1) | 0.989966 | 0.989997 |
| Table 6.1 (set 2) | 0.999983 | 0.999993 | Table 6.1 (set 2) | 0.999833 | 0.999906 |
| Table 6.1 (set 3) | 0.999923 | 0.999996 | Table 6.1 (set 3) | 0.999887 | 0.999976 |
| Table 6.1 (set 4) | 0.999968 | 0.999995 | Table 6.1 (set 4) | 0.999857 | 0.999989 |
| Table 6.1 (set 5) | 0.999989 | 0.999999 | (set 5) | 0.999936 | 0.999953 |

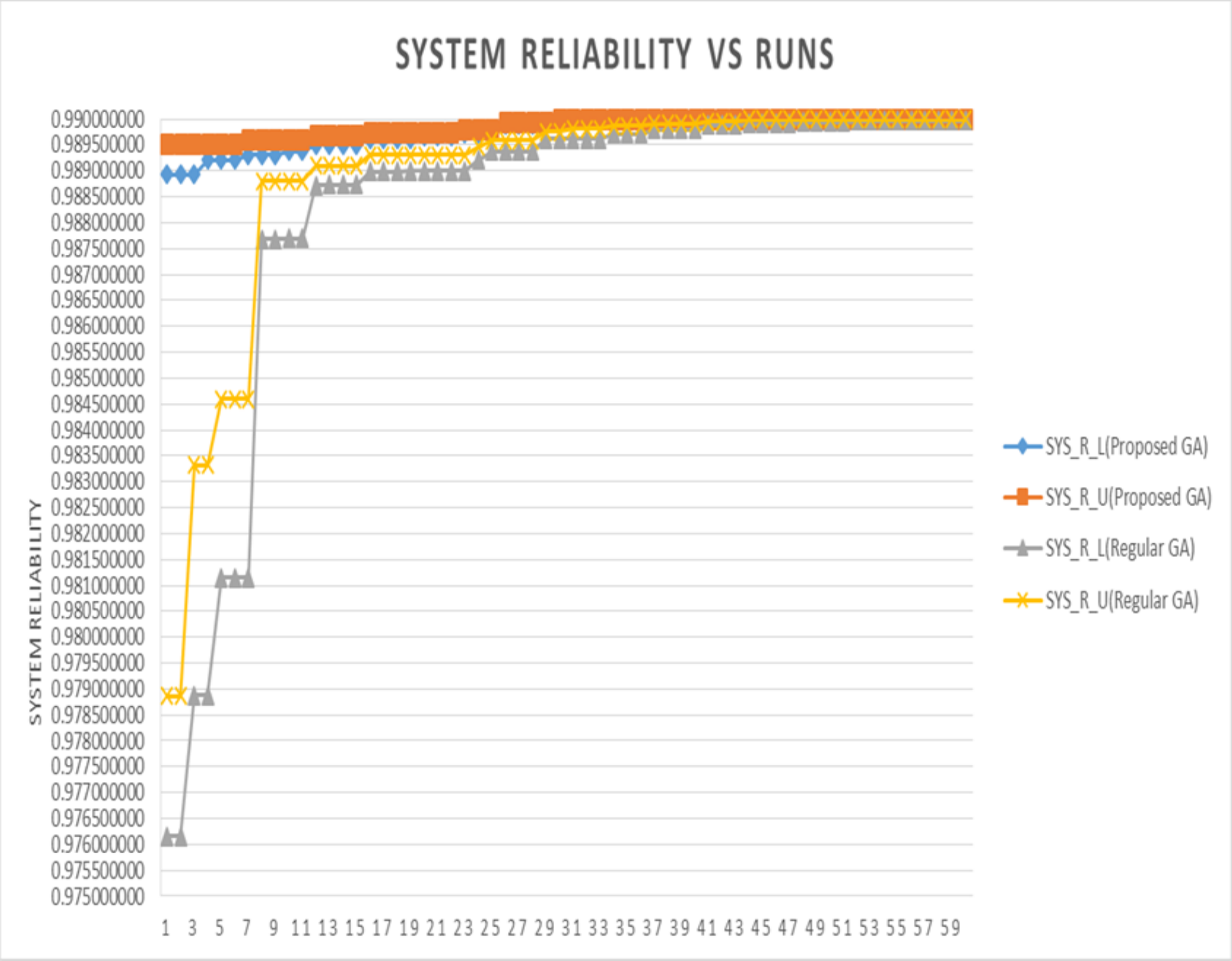

**Figure 1.9 Solution evolution for evaluation example-two**

**Table 11 Optimal solution for the evaluation example-two with 6OICs and 10 functions using GA**

| **Proposed GA** | **Table 6.1(set 1)** | **Table 6.1(set 2)** |
|---|---|---|
| **X** | 1 0 0 0 0 0 1 1 1 0<br>0 0 0 0 0 0 1 1 0 0<br>0 0 0 0 0 0 0 0 0 1<br>0 1 0 0 0 0 0 0 0 0<br>0 1 0 0 1 1 0 1 1 0<br>0 0 0 0 1 0 0 0 0 1 | 0 1 0 0 0 1 0 0 1 0<br>0 1 0 0 0 0 1 1 0 0<br>0 0 1 0 1 0 1 0 0 0<br>0 0 0 0 0 0 0 0 0 1<br>1 0 0 0 1 1 0 0 1 0<br>0 0 0 1 0 0 0 0 0 0 |
| U:<br>**A**<br>OICs | 3 3 3 1 3 3 3 3 4 4<br><br>6 | 5 2 4 2 3 2 4 3 3 2<br><br>6 |
| System Reliability<br>Lower/Upper | 0.989966 / 0.989997 | 0.999968/0.999984 |
| Cost (in cycles) | 2500 | 1900 |
| Regular GA | | |
| X | 1 0 0 0 0 1 1 1 0 0<br>0 1 1 0 1 0 1 0 0 0<br>0 0 0 0 0 1 0 0 1 0<br>1 0 0 1 0 0 0 0 0 0<br>1 1 0 0 0 0 0 0 0 0<br>1 0 0 0 1 1 0 1 1 0 | 0 0 0 1 0 1 1 0 0 0<br>0 0 0 1 0 0 0 0 0 1<br>1 0 0 0 1 1 0 0 1 0<br>0 0 0 1 0 1 0 0 0 0<br>1 0 1 0 0 0 0 0 0 0<br>0 0 0 0 0 0 0 0 0 0 |
| U:<br>A<br>OICS | 4 3 3 3 3 3 3 3 4 4<br><br>6 | 4 4 5 3 6 5 3 5 3 3<br><br>6 |
| System Reliability<br>Lower/Upper | 0.989876 / 0.989991 | 0.999006 / 0.999646 |
| Cost(in cycles), | 2500 | 1900 |

**Table 11 (Continued)**

| **Proposed GA** | **Table 6.1(set 3)** | **Table 6.1(set 4)** | **Table 6.1(set 5)** |
|---|---|---|---|
| **X** | 0 0 0 0 0 0 1 1 0 0<br>0 1 0 0 0 0 0 0 0 0<br>0 0 0 0 0 0 1 1 0 0<br>0 0 0 0 0 0 1 0 0 0<br>0 0 0 1 0 0 0 0 1 0<br>0 0 0 0 0 1 0 1 1 0 | 0 0 0 1 0 1 0 1 0 0<br>0 1 0 0 0 0 0 0 0 0<br>1 0 0 1 0 0 0 0 1 0<br>1 0 0 0 1 0 0 0 0 0<br>0 0 0 0 0 0 0 1 0 0<br>0 0 0 1 1 0 0 1 0 1 | 0 0 0 0 0 0 1 1 0 1<br>0 0 0 0 0 1 1 0 0 0<br>0 1 0 0 0 0 0 0 0 0<br>0 0 1 0 0 1 0 0 0 1<br>0 0 0 0 0 1 0 0 0 1<br>0 0 0 0 1 0 0 0 0 0 |
| U:<br>**A**<br>OICs | 3 3 2 2 3 3 3 2 3 3<br><br>6 | 4 3 3 3 3 2 3 3 3 4<br><br>6 | 2 3 3 3 3 3 4 3 3 3<br><br>6 |
| System Reliability Lower/Upper | 0.999602/0.999968 | 0.999903 / 0.999945. | 0.999977 / 0.999997 |
| Cost (in cycles) | 1900 | 2150 | 2000 |
| Regular GA | | | |
| X | 0 0 0 0 0 0 0 1 0 0<br>0 0 0 0 0 0 0 0 0 0<br>1 0 0 1 0 0 1 1 1 0<br>1 0 0 0 1 0 0 0 0 0<br>0 0 0 0 0 0 0 1 0 0<br>0 0 0 1 1 0 0 1 0 1 | 0 0 0 0 0 0 0 0 0 0<br>0 0 1 1 0 0 0 0 0 0<br>0 0 0 0 0 0 0 0 0 0<br>0 0 0 0 0 1 0 0 0 0<br>1 0 0 0 0 0 0 0 0 0<br>0 0 0 0 0 0 0 1 1 1 | 1 1 1 0 0 0 0 1 0 1<br>0 0 0 0 1 0 0 0 0 1<br>0 0 0 1 0 0 0 0 0 0<br>1 0 1 1 0 0 0 0 0 0<br>1 1 1 1 0 1 0 0 1 0<br>0 0 0 1 1 1 1 1 0 0 |
| U:<br>A<br>OICS | 3 3 2 2 4 3 3 4 3 3<br><br>6 | 4 3 3 3 3 2 3 3 3 3 3<br><br>6 | 3 3 3 4 3 3 4 3 3 3<br><br>6 |
| System Reliability Lower/Upper | 0.999791 / 0.999991 | 0.999860 / 0.999917 | 0.999851/0.999952 |
| Cost(in cycles), | 2350 | 2200 | 2350 |

## 8.4 Sensitivity Analysis

The sensitivity analysis is performed to study the impact of GA parameters on the objective function. The analysis also helps to study the stability of the system when impreciseness is addressed with interval valued numbers in the objective function. GA parameter values are also selected using this analysis. Here the stability of the proposed GA for all interval numbers (Table 5) are illustrated using graphs in the Figure 1.10 – 1.29 with respect to GA parameters (P_SIZE, P_MUTAT, P_CROSS, and M_GENER). It is observed from the Figure 1.10 -1.29 proposed GA is steady and unchanging, thereby providing quality solutions for the proposed RAP formulation. A few important observations are enumerated below with respect to GA parameter selection and their consistency.

(a) System Reliability (Lower/Upper) in Figures 1.10, 1.14, 1.18, 1.22 and 1.26 is monotonically increasing with generations. The parameter M_GENER = 200 is a suitable value that reflects the stability of system reliability (lower/upper) for all interval numbers stated in Table 5.

(b) P_SIZE (population size) is varied with respect to the system reliability (lower/upper) in Figures 1.11, 1.15, 1.19, 1.23, and 1.27. Notably, the stochastic selection is adopted in the proposed GA. Even then, system reliability (lower/upper) is increasing proportionally with respect to a population size and stabilizes at P_SIZE =100. P_SIZE does not impact the optimal solution of our proposed GA with population size of 100 or above.

(c) P_CROSS: the impact of crossover on the system reliability is shown in Figures 1.12, 1.16, 1.20, 1.24, and 1.28. It is factual from these figures, that an increase in crossover rate does not have any impact on the system reliability (lower/upper). An increase in crossover rate does not result in an abnormal change in the system reliability values.

(d) P_MUTAT: A higher mutation level generates greater genetic assortment and sometimes does have an impact on the system. Figures 1.13, 1.17, 1.21, and 1.29 reflect stability in the system reliability. Higher mutation rate with elitism in the proposed GA sets the population for better solution evolution in the due course with recovery of lost solutions in the earlier generations. Both P_CROSS and P_MUTAT rate chosen for the RAP are suitable to sustain stability over a large range.

(e) Choice of the class of interval numbers for component reliabilities is not relevant because, there can be only one optimal system with M and N. Selection of interval valued numbers does not influence the optimal solution of the system as discerned from sensitive analysis. The system reliability monotonically increases with the number of components and functions deployed. Thus, the optimal solution by the proposed GA is not affected. The stability of the system is ascertained.

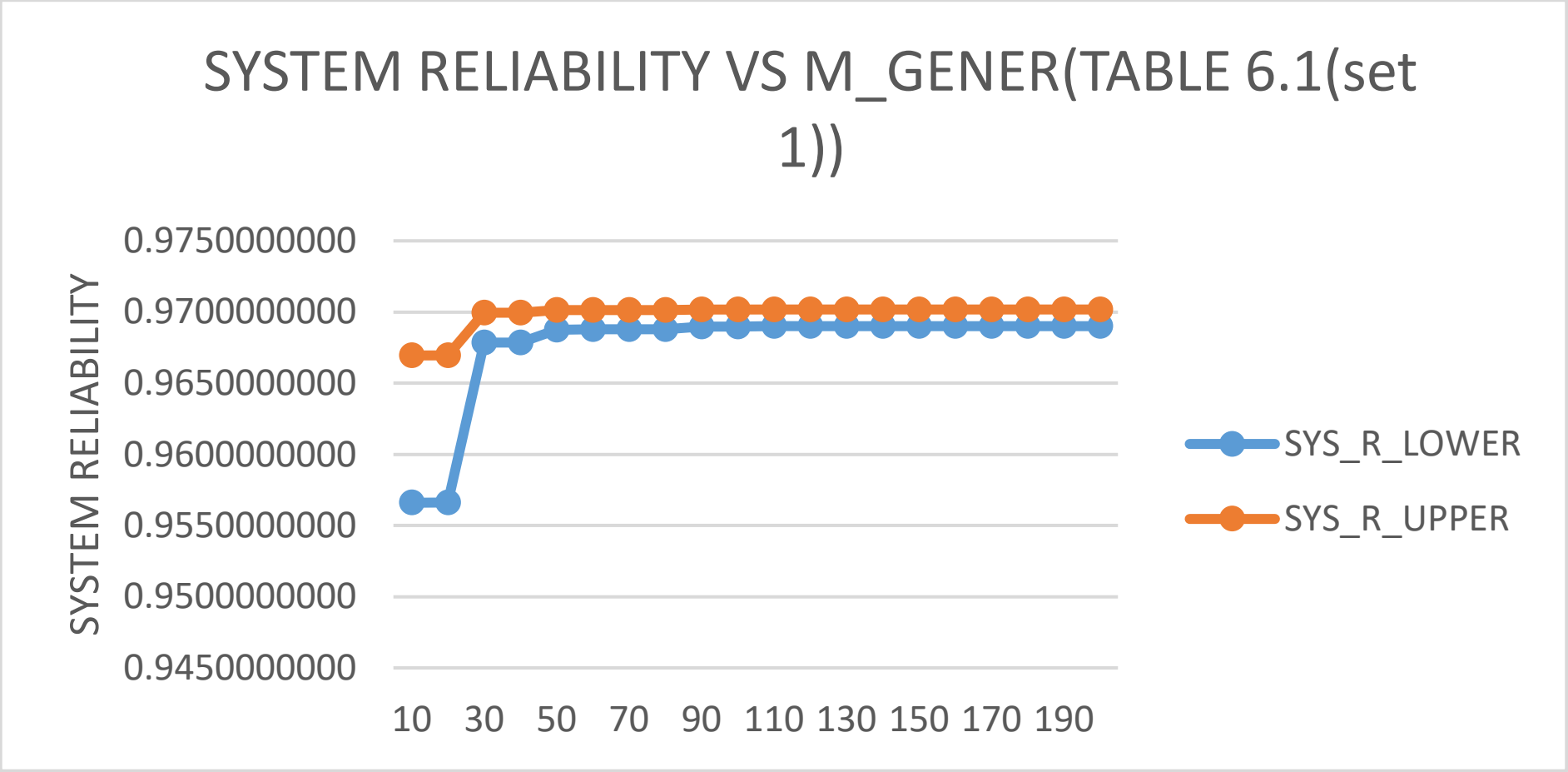

**Figure 1.10 System Reliability Vs Generations for Interval numbers set 1 in Table 5.**

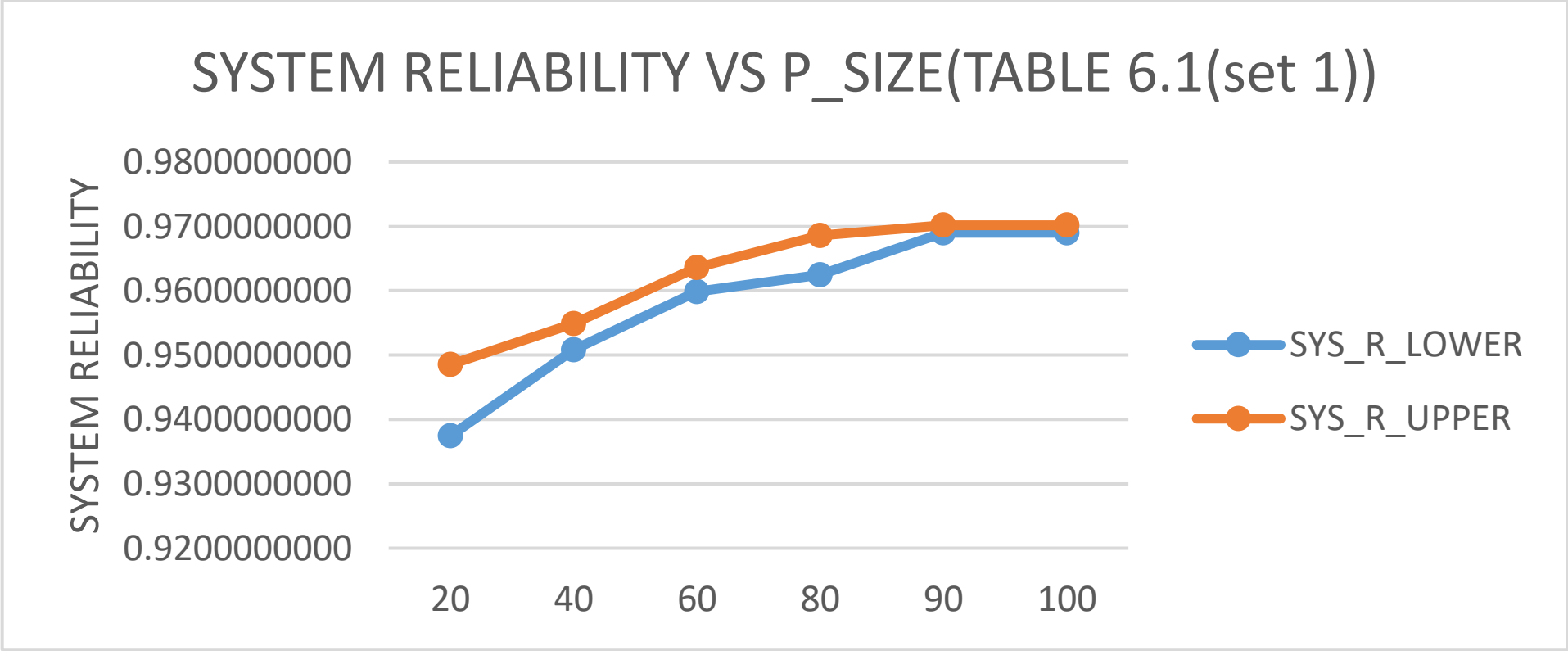

**Figure 1.11 System Reliability Vs Population Size for Interval numbers set 1 in Table 5**

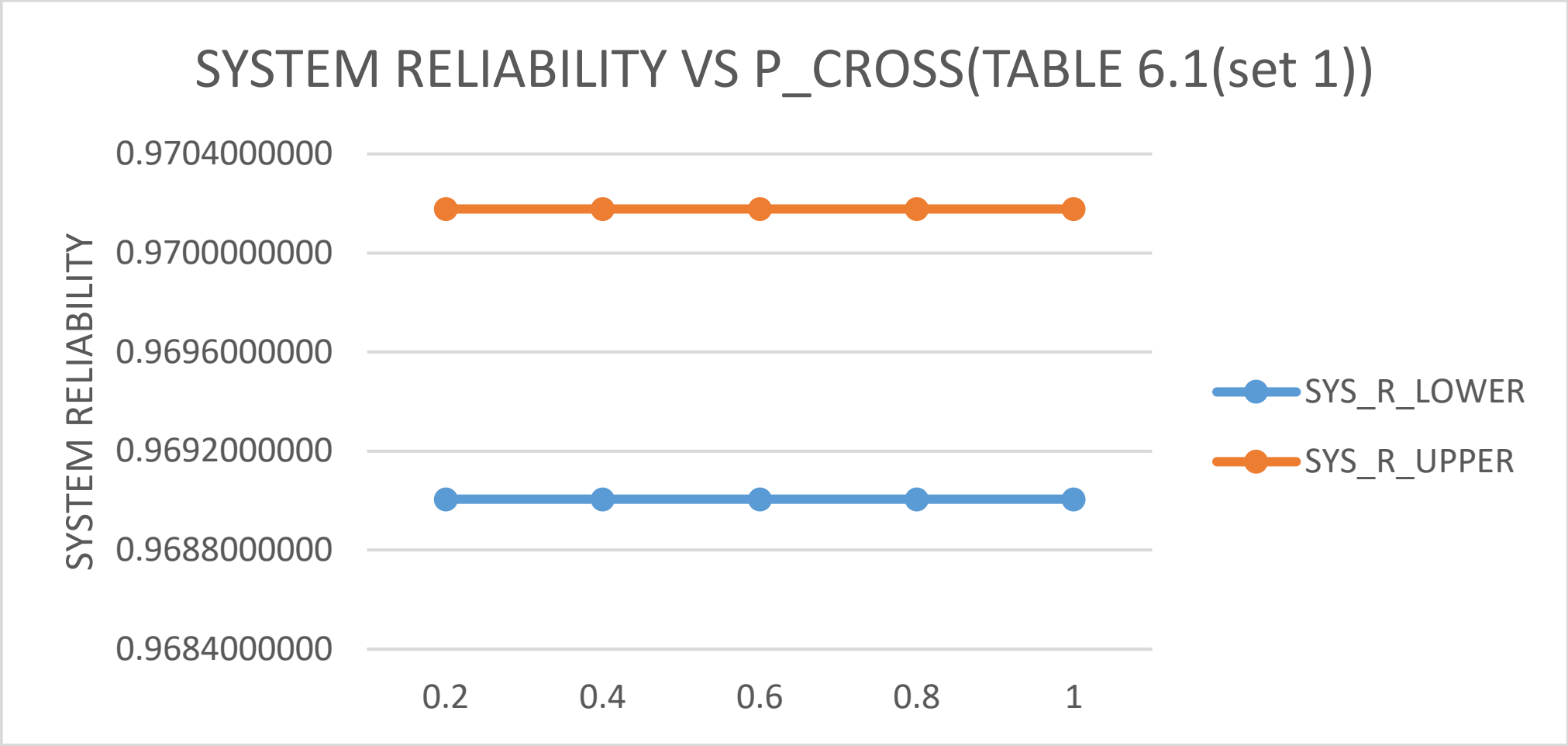

**Figure 1.12 System Reliability Vs Probability of Crossover for Interval numbers set 1 in Table 5**

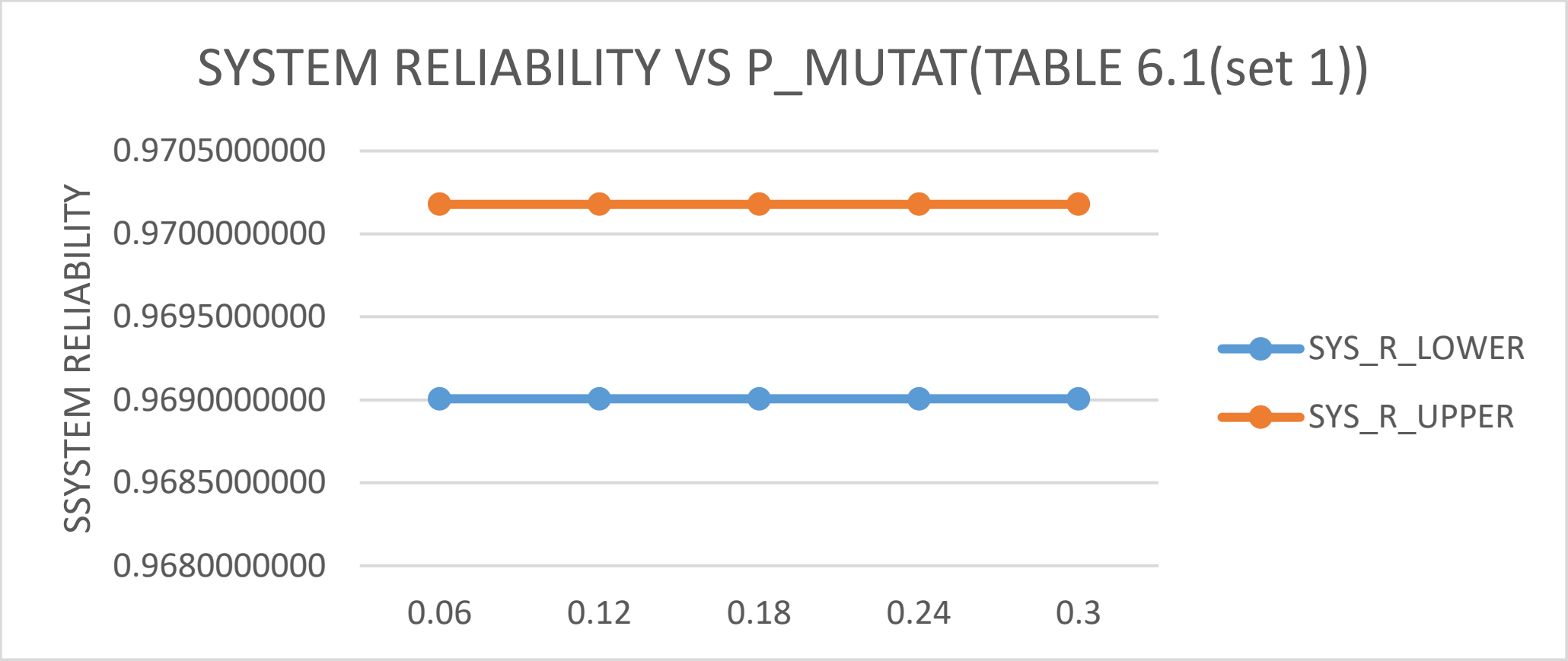

**Figure 1.13 System Reliability Vs Probability of Mutation for Interval numbers set 1 in Table 5**

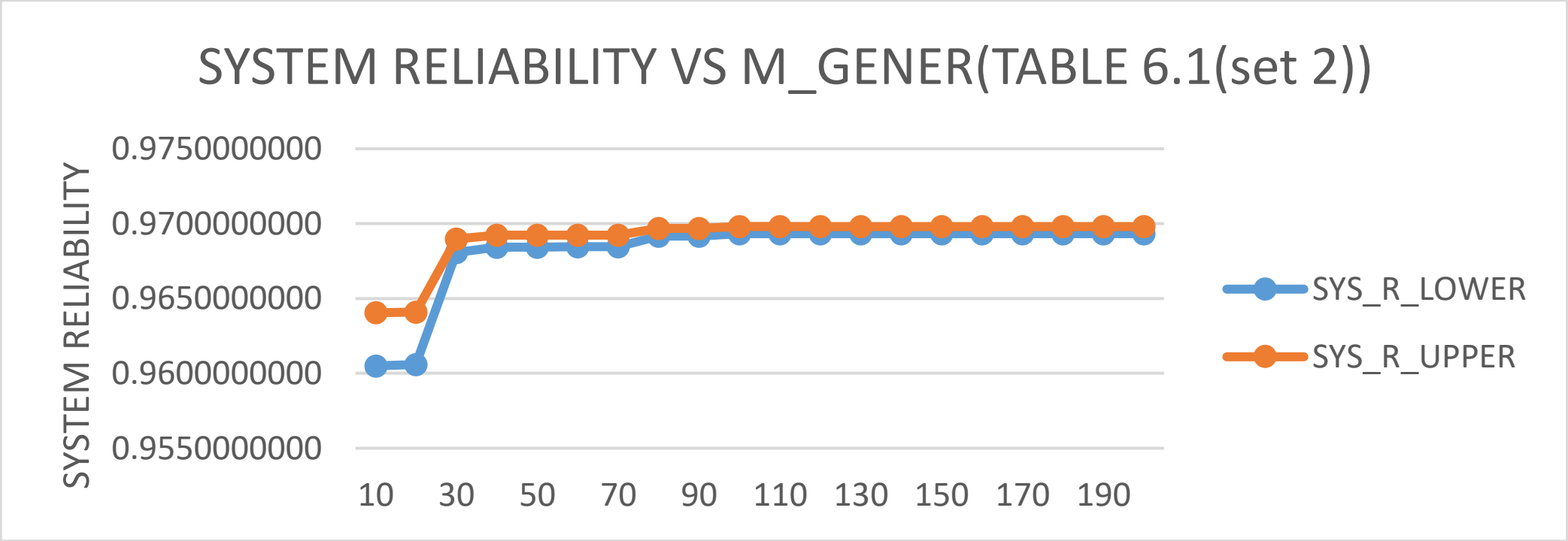

**Figure 1.14 System Reliability Vs Generations for Interval numbers set 2 in Table 5**

**Present version is version 1.0. Manuscript is under Review. Modified or extended version will appear in the Journal.**

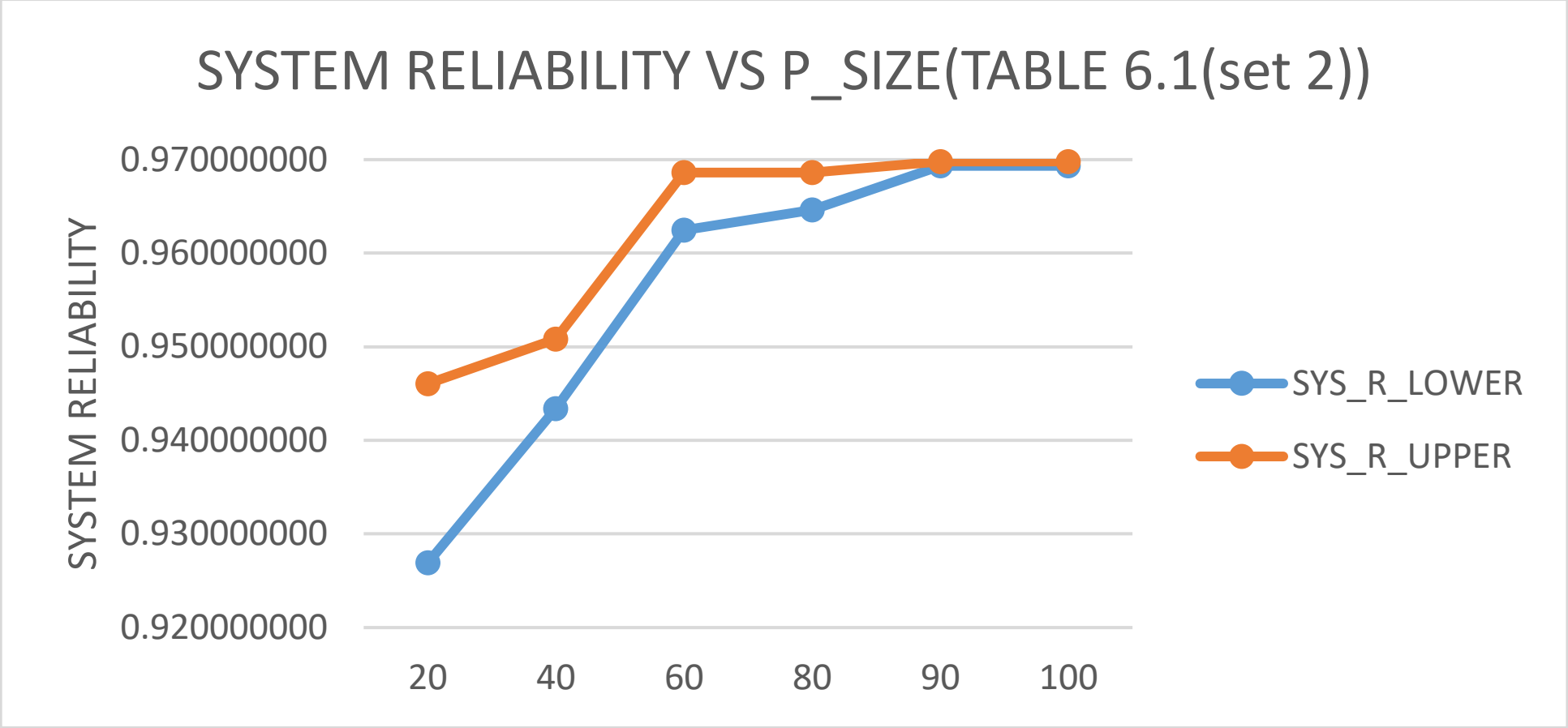

**Figure 1.15 System Reliability Vs Population Size for Interval numbers set 2 in Table 5**

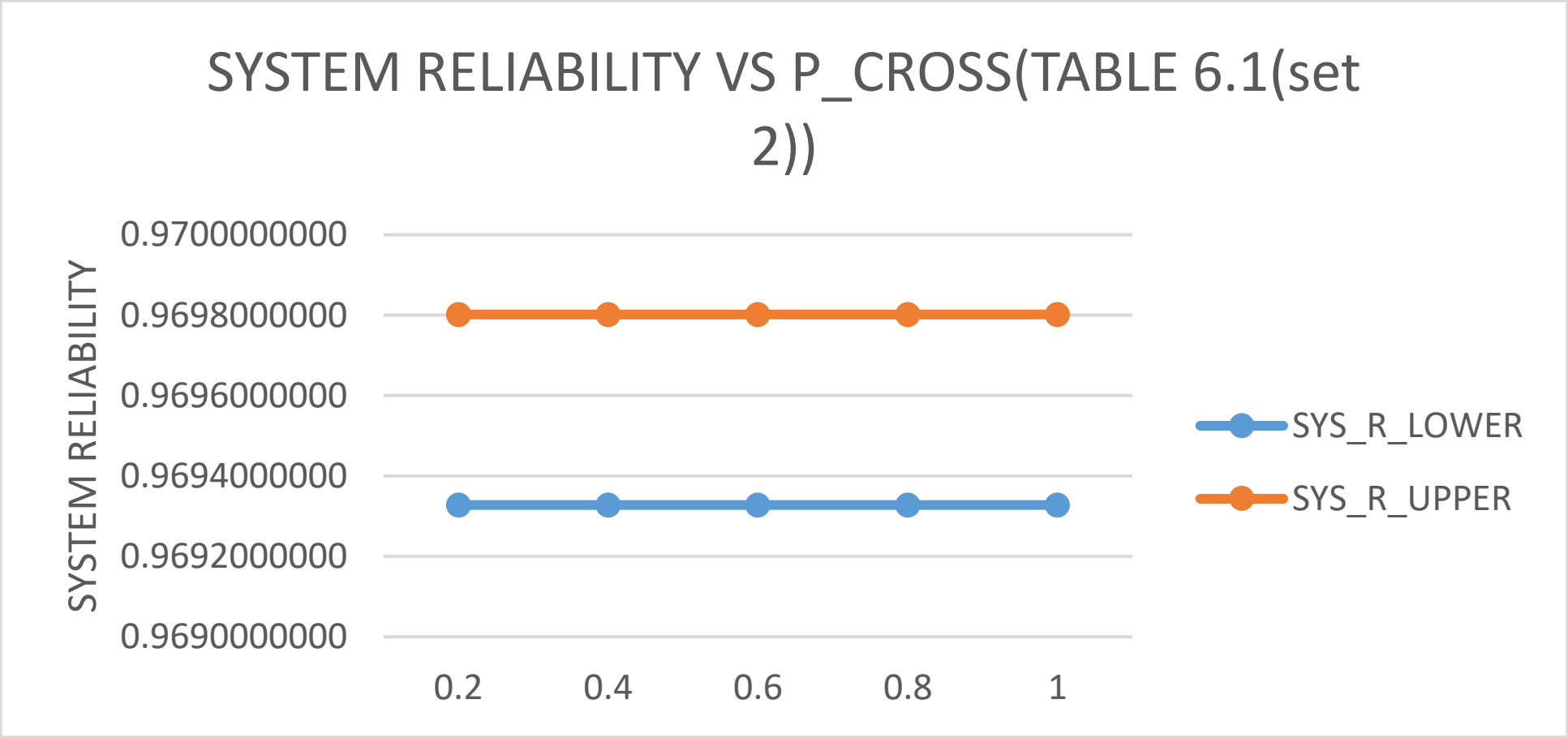

**Figure 1.16  System Reliability Vs Probability of Crossover for Interval numbers set 2 in Table 5**

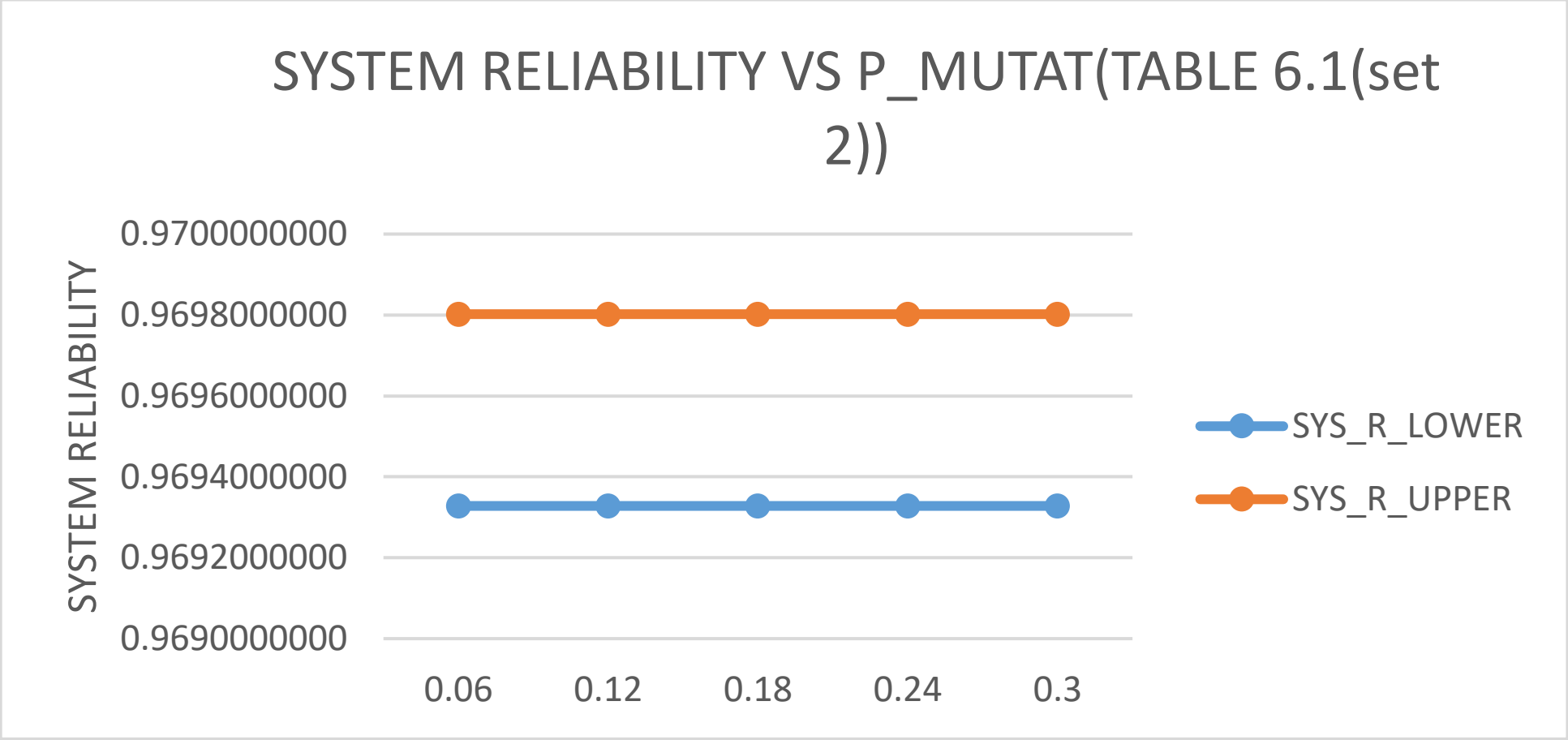

**Figure 1.17 System Reliability Vs Probability of Mutation for Interval numbers set 2 in Table 5**

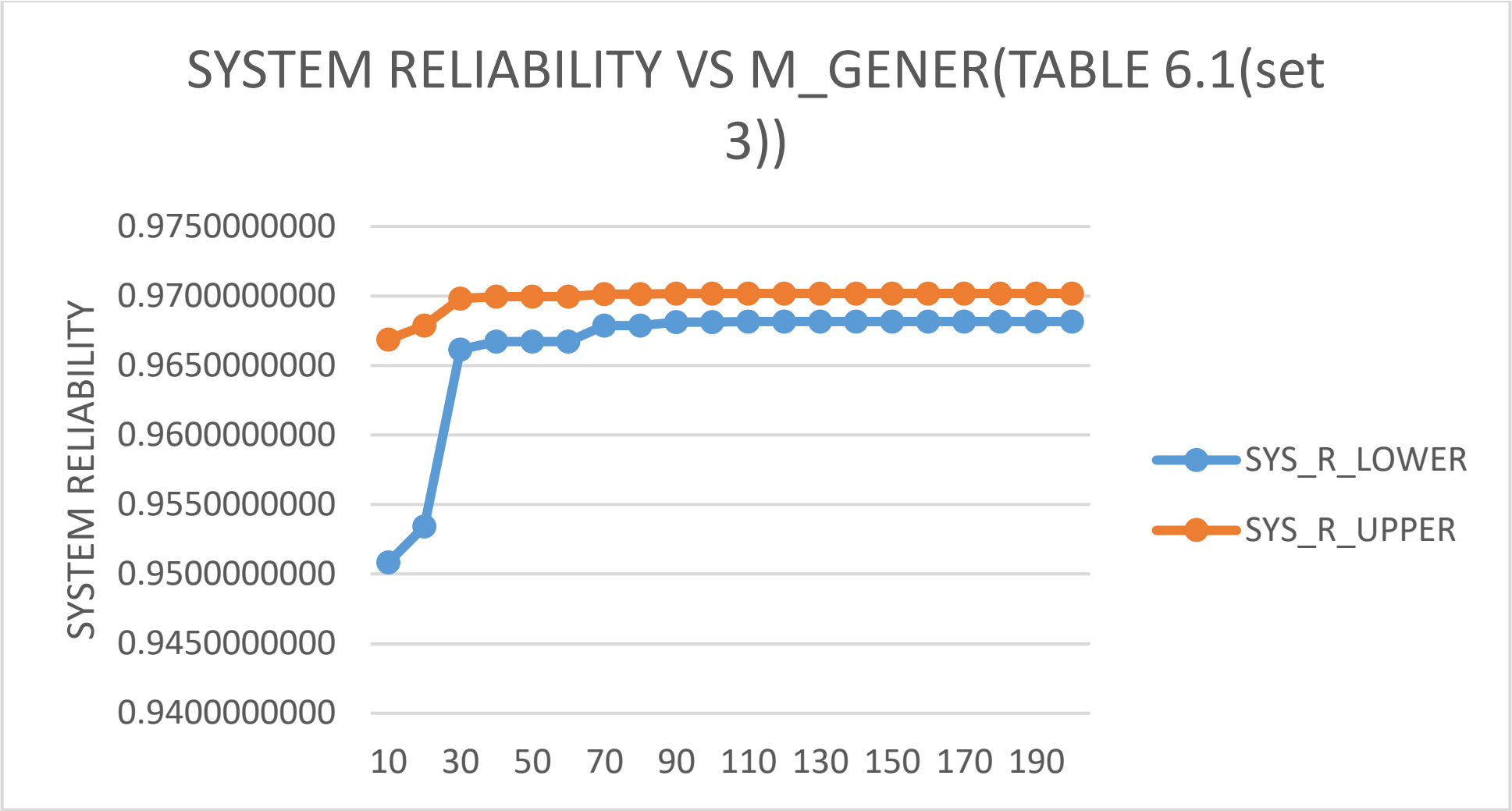

**Figure 1.18 System Reliability Vs Generations for Interval numbers set 3 in Table 5**

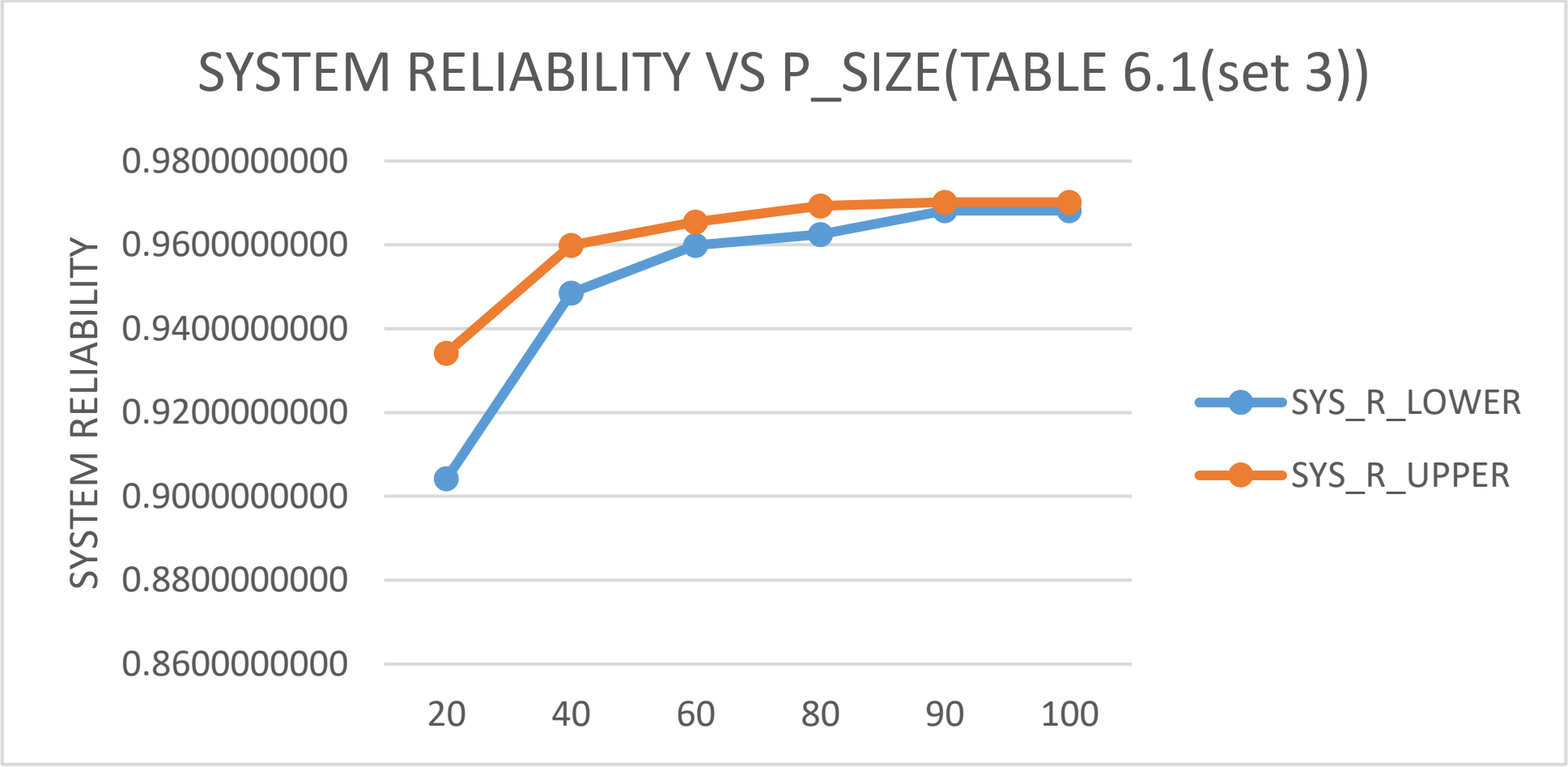

**Figure 1.19 System Reliability Vs Population Size for Interval numbers set 3 in Table 5**

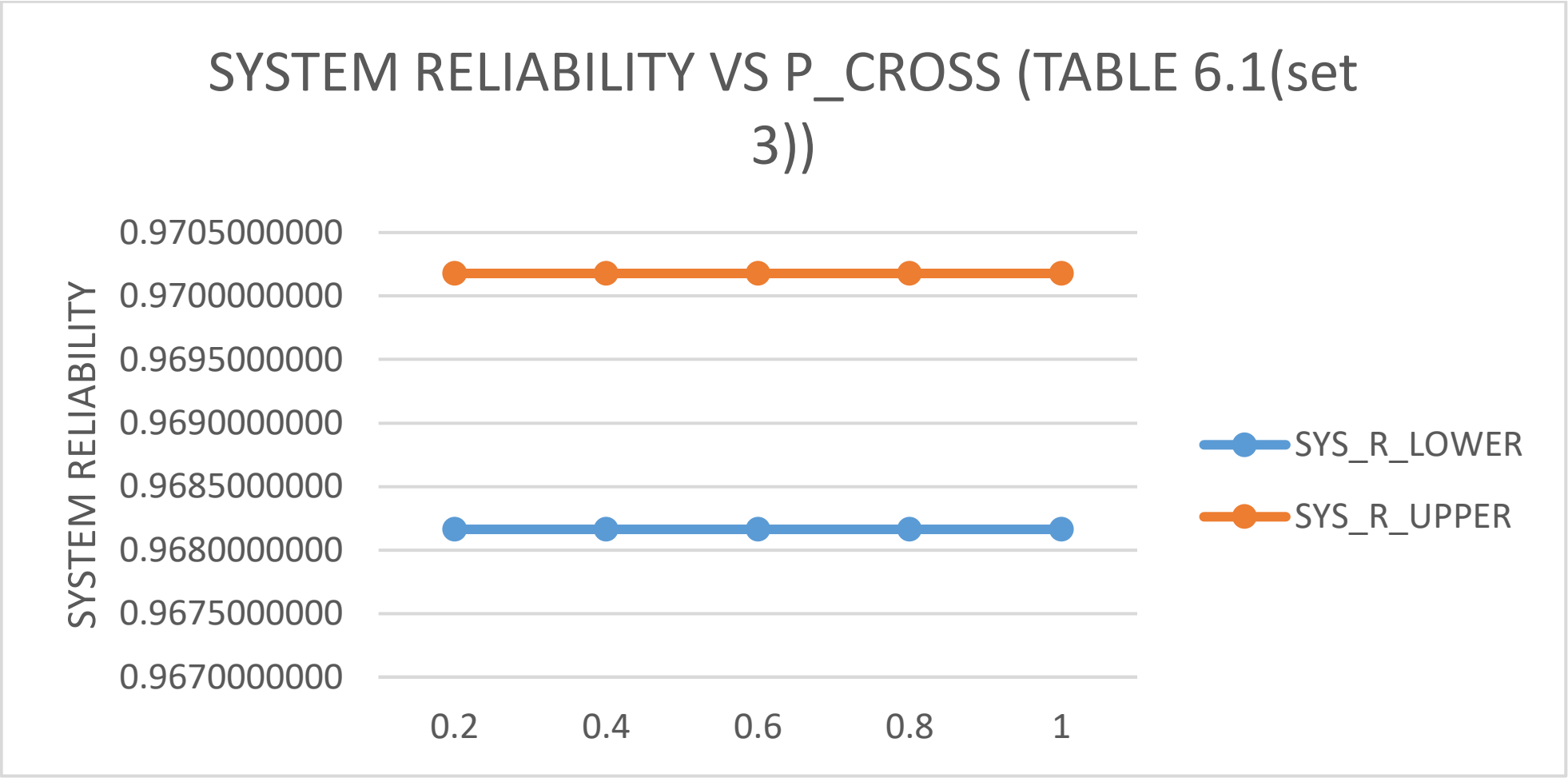

**Figure 1.20 System Reliability Vs Probability of Crossover for Interval numbers set 3 in Table 5**

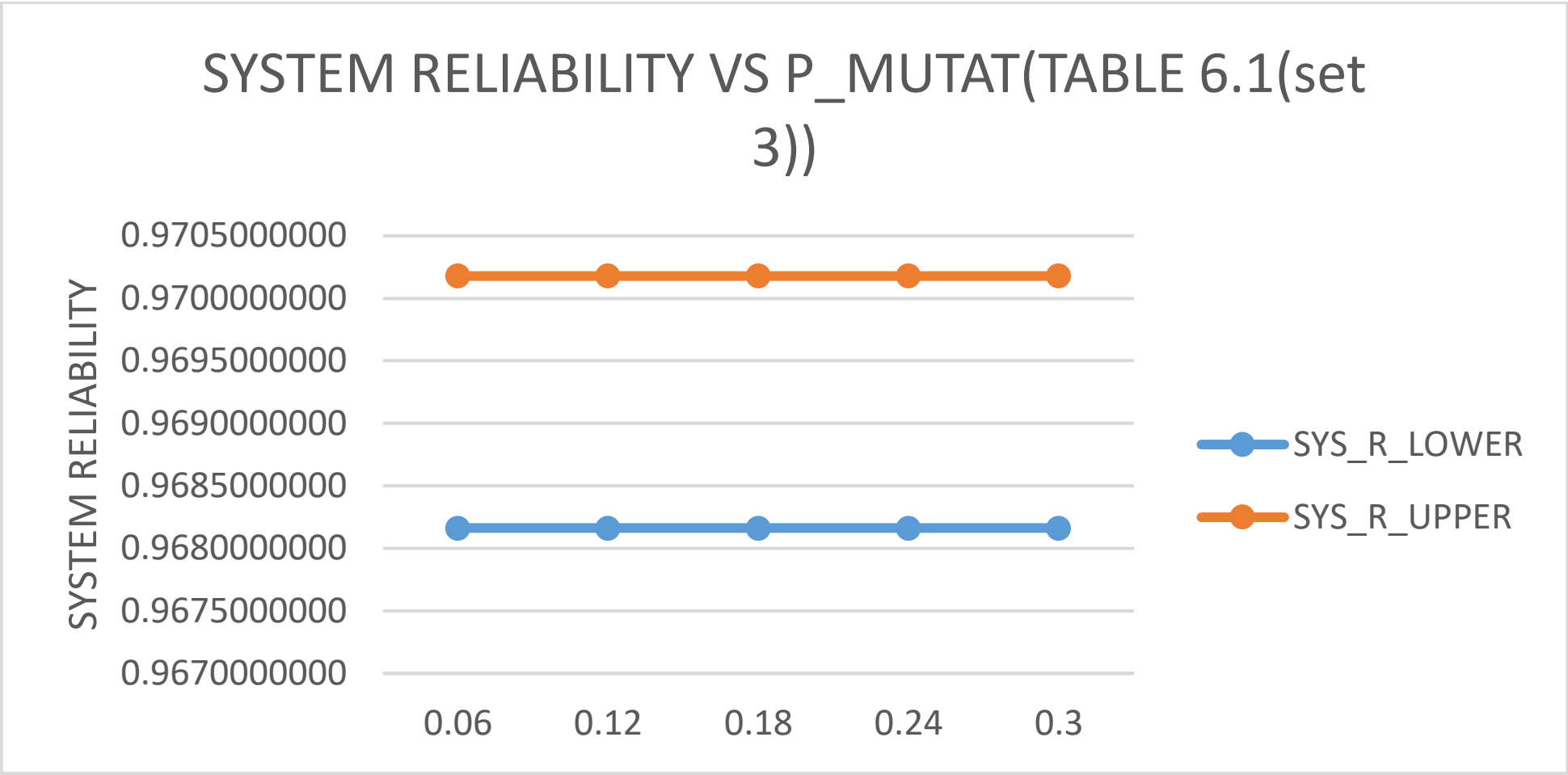

**Figure 1.21 System Reliability Vs Probability of Mutation for Interval numbers set 3 in Table 5**

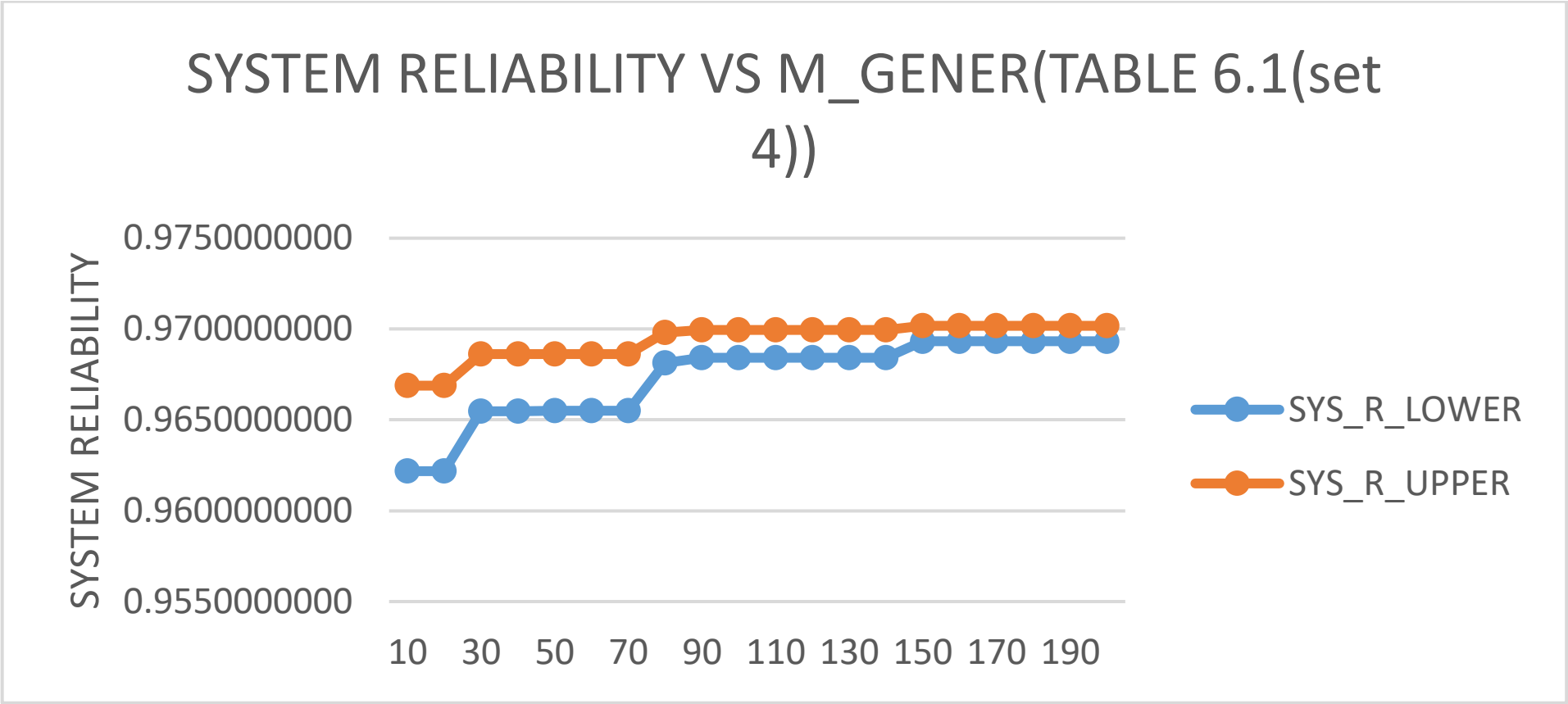

**Figure 1.22 System Reliability Vs Generations for Interval numbers set 4 in Table 5**

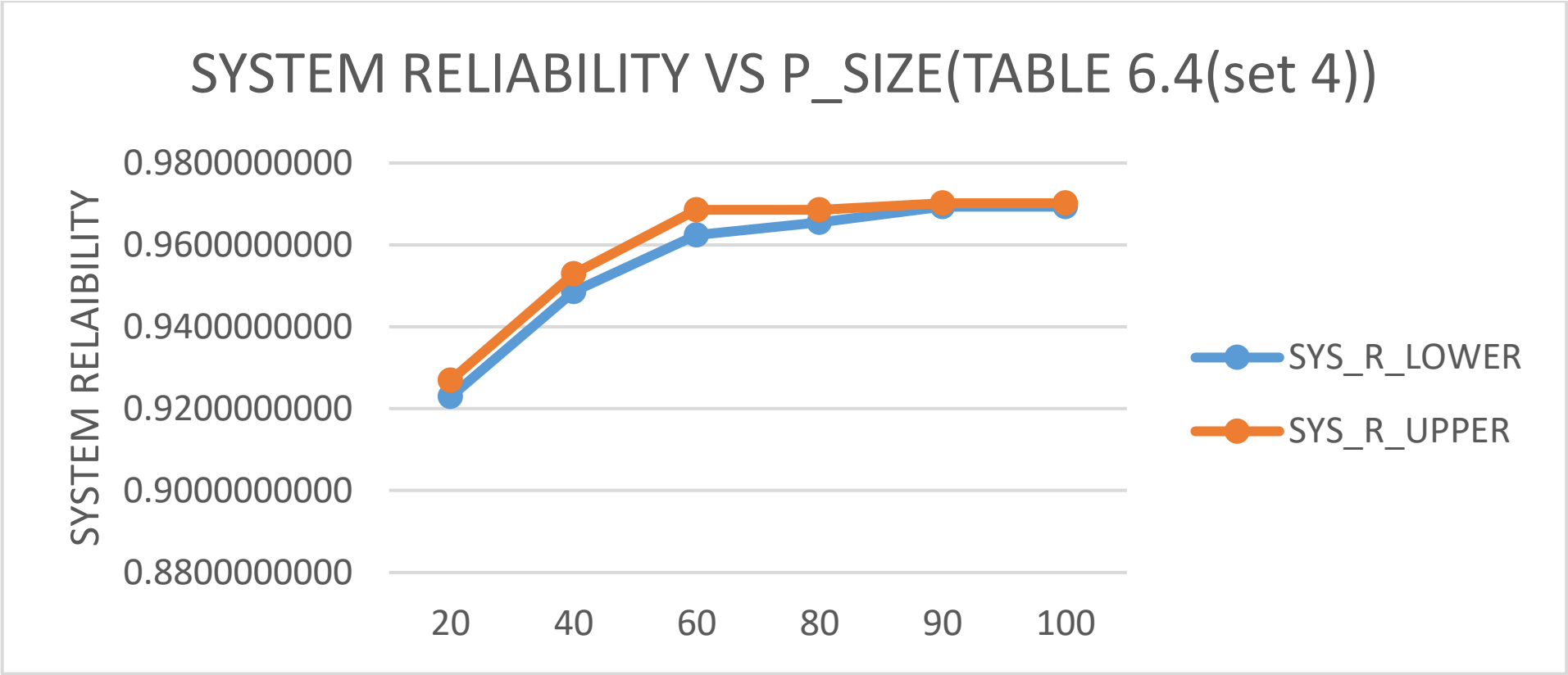

**Figure 1.23 System Reliability Vs Population Size for Interval numbers set 4 in Table 5**

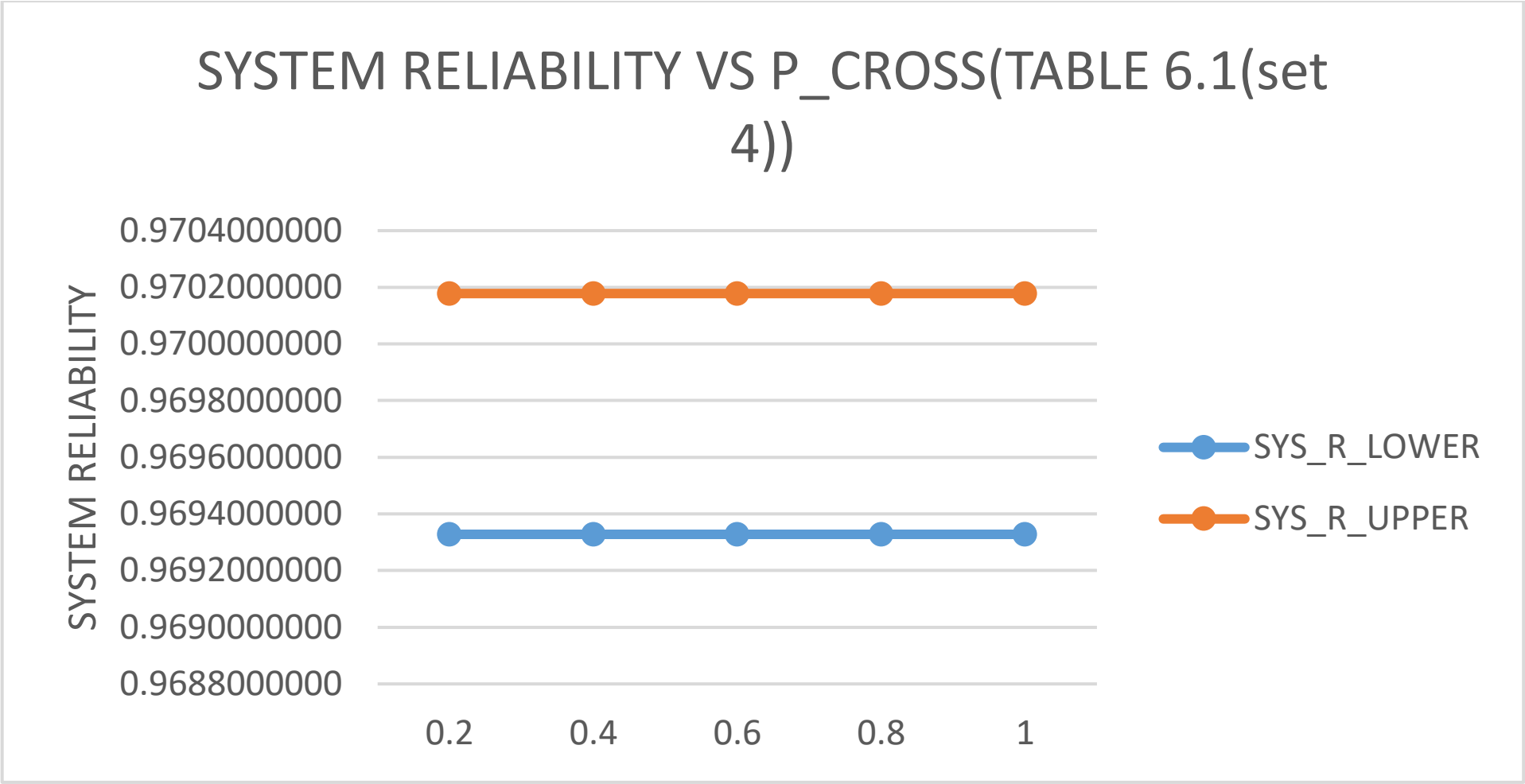

**Figure 1.24 System Reliability Vs Probability of Crossover for Interval numbers set 4 in Table 5**

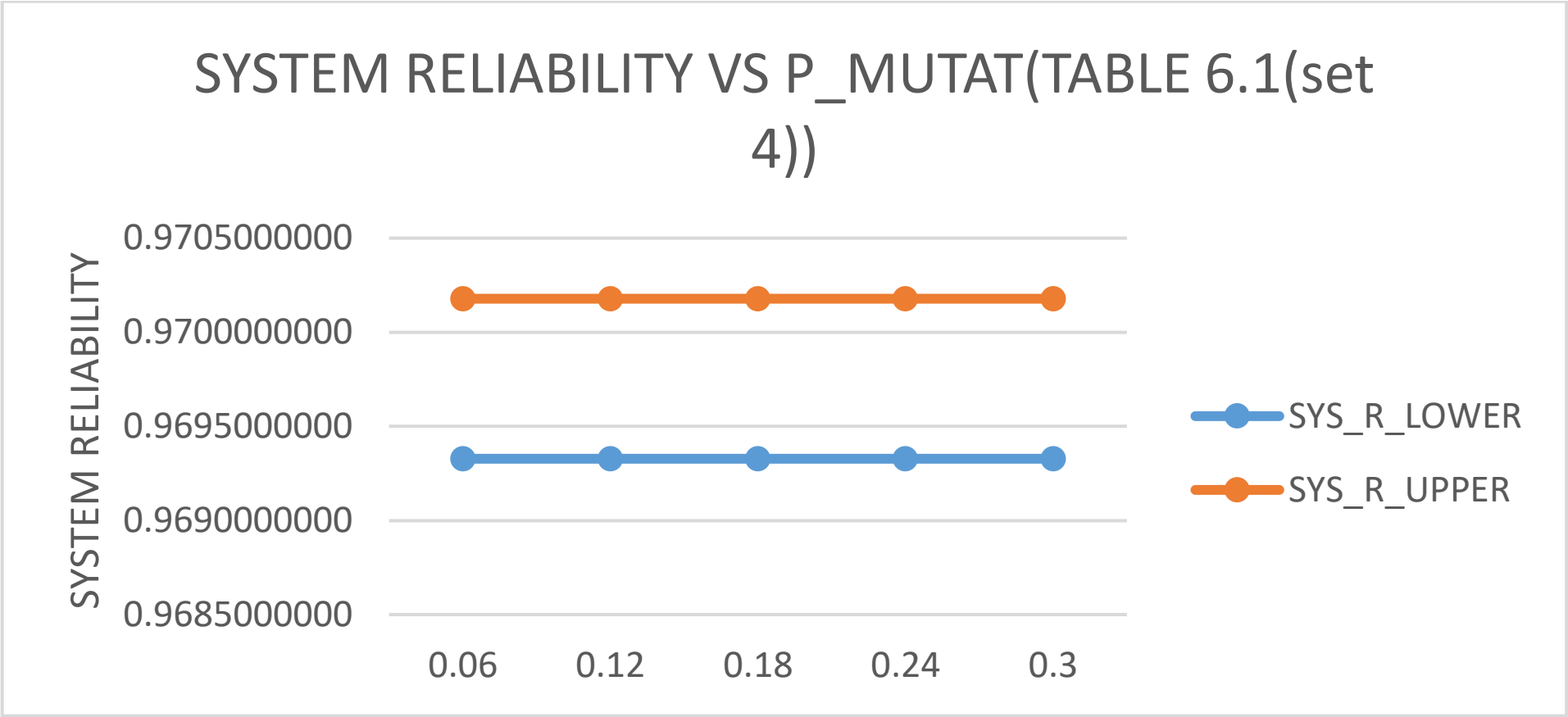

**Figure 1.25 System Reliability Vs Probability of Mutation for Interval numbers set 4 in Table 5**

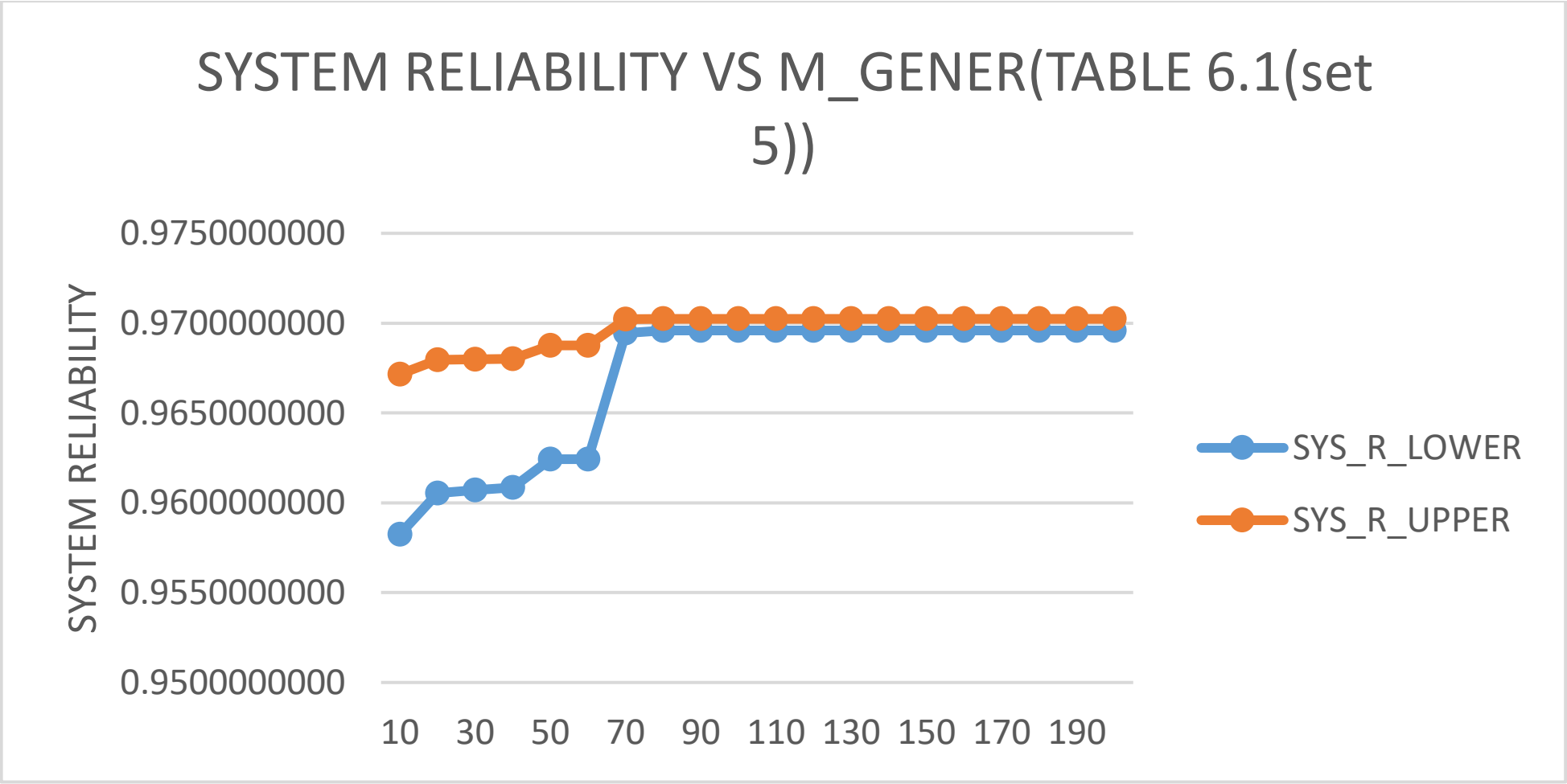

**Figure 1.26 System Reliability Vs Generations for Interval numbers set 5 in Table 5**

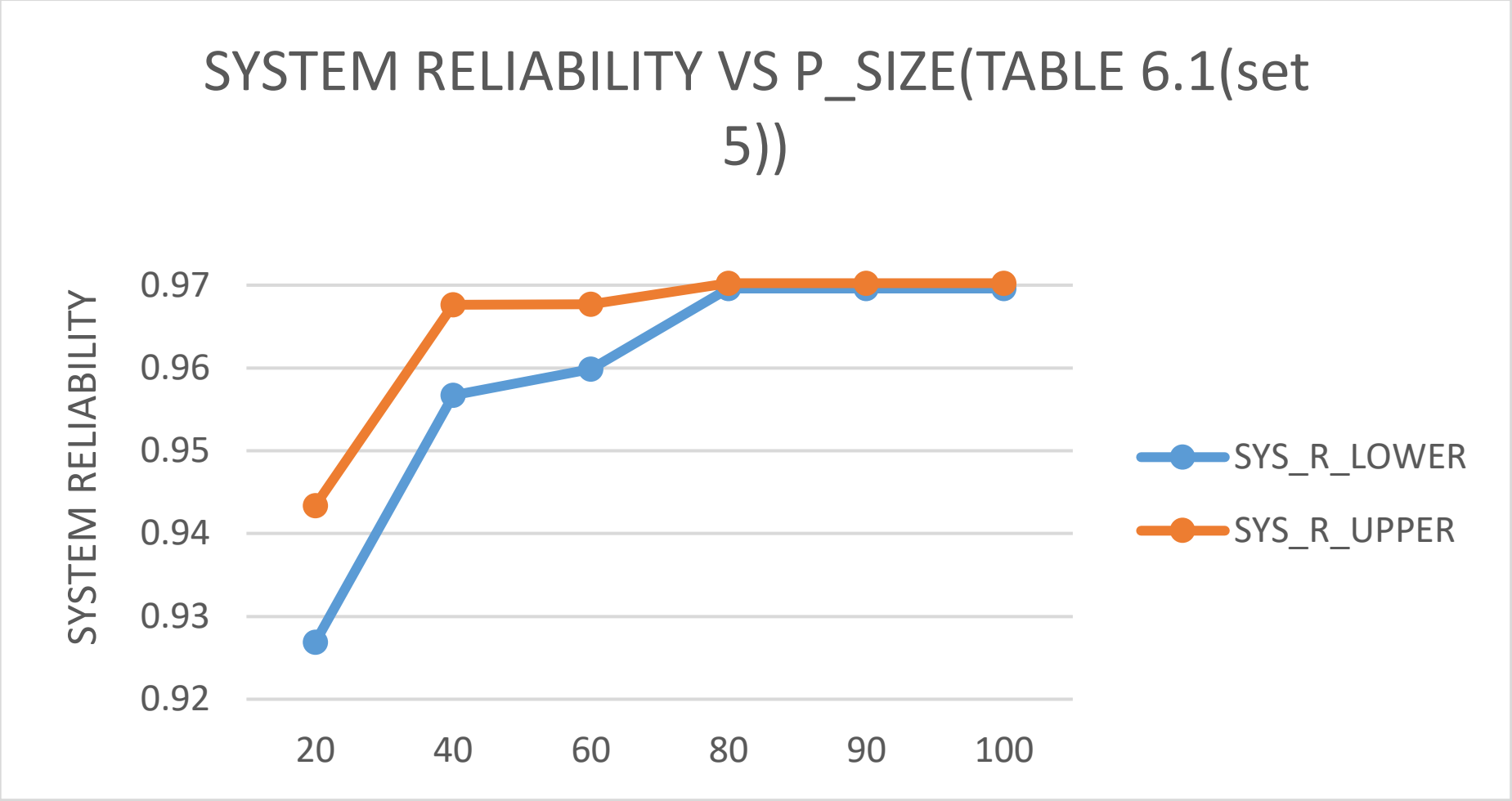

**Figure 1.27 System Reliability Vs Population Size for Interval numbers set 5 in Table 5**

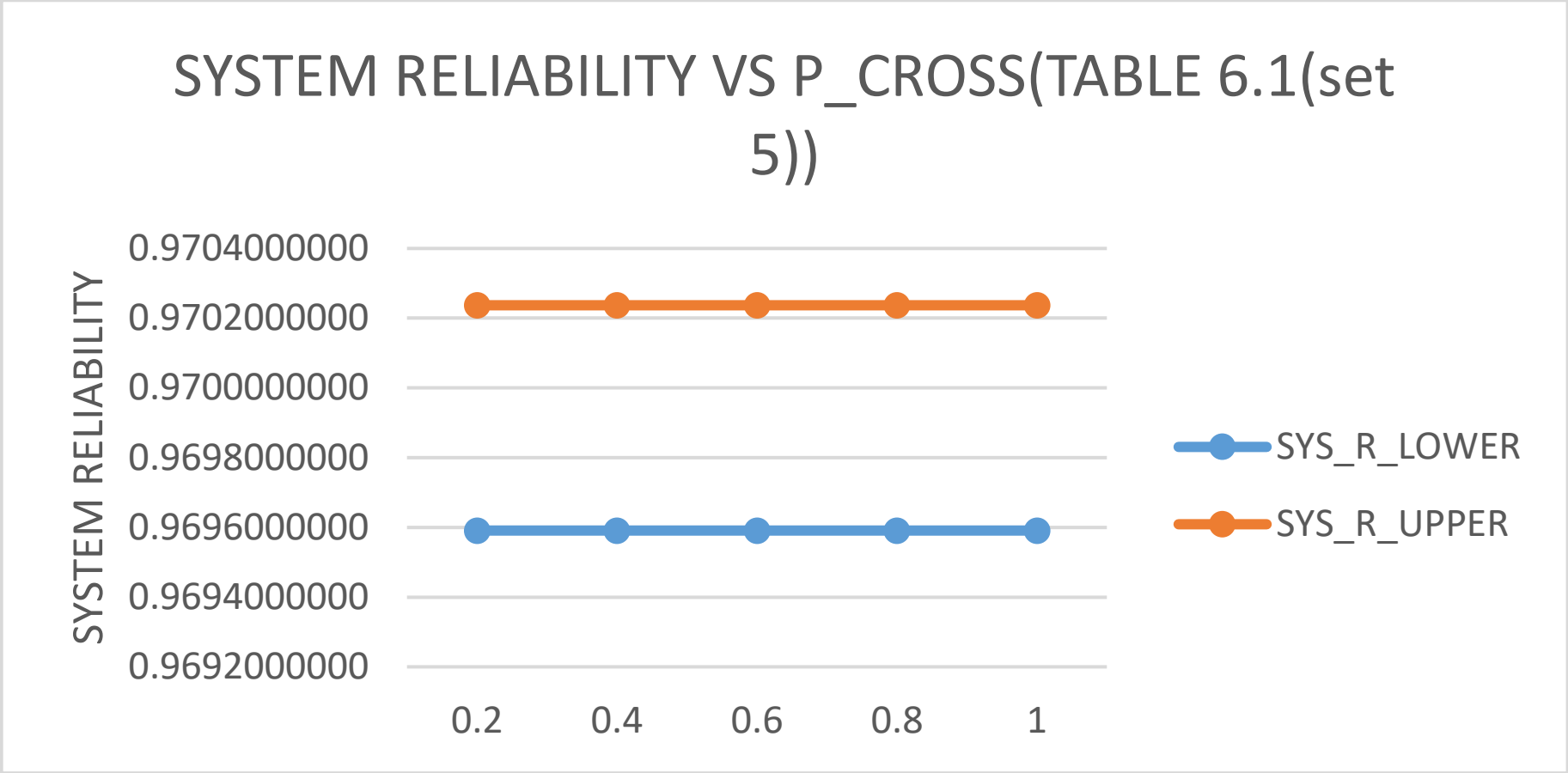

**Figure 1.28 System Reliability Vs Probability of Crossover for Interval numbers set 5 in Table 5**

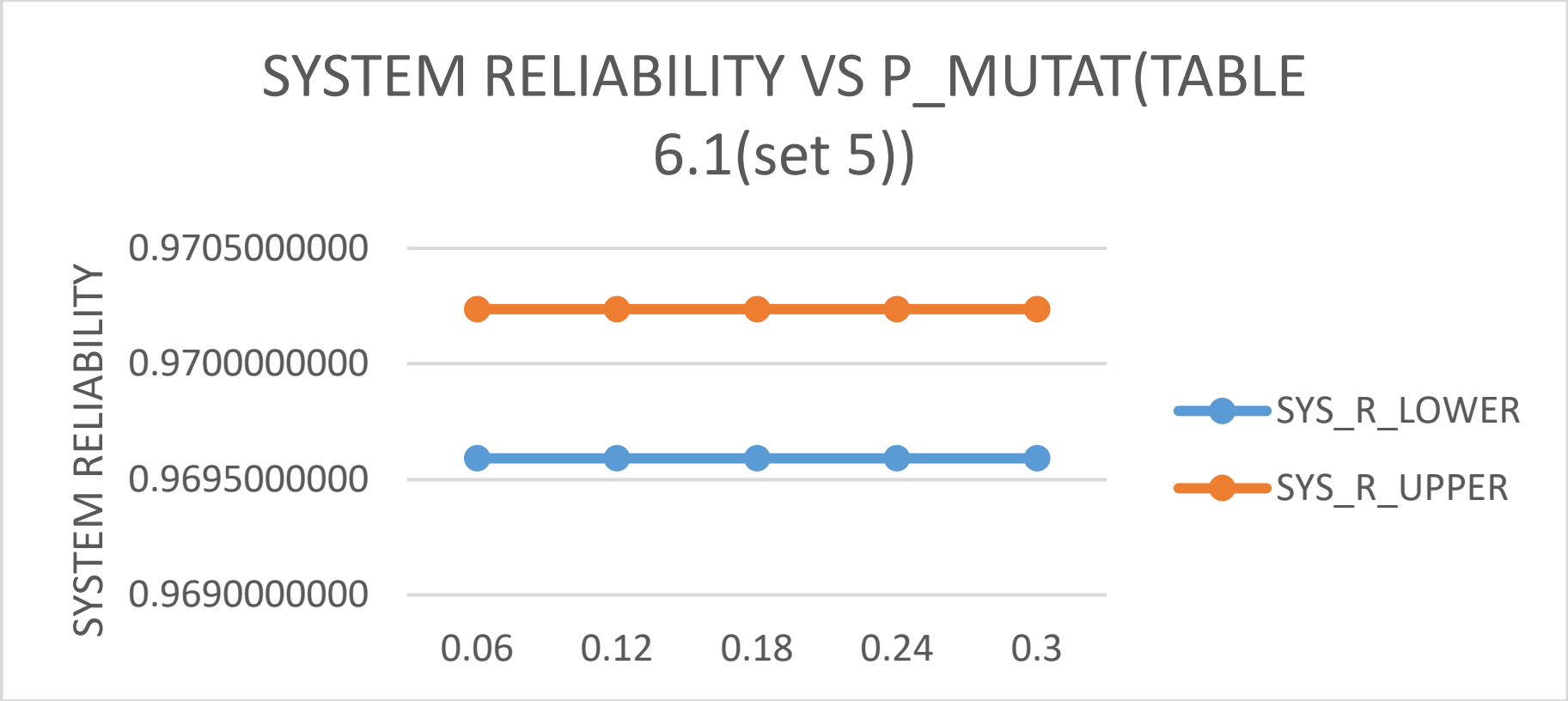

**Figure 1.29 System Reliability Vs Probability of Mutation for Interval numbers set 5 in Table 5**

# 9. Particle Swarm Optimization for Interval RAP

The Particle Swarm Optimization (PSO) is an effective alternative to solve large non-linear optimization problems. PSO is proposed for Interval RAP with the objective function given in the Equation (47) as the fitness function. The flocking of birds is used for conceptualising the search framework in PSO. This conceptual framework assists in finding local and global solution for the optimization problem unlike GA where only global search is performed. In PSO, a swarm of particles, each represents a candidate solution for a given optimization problem is maintained. All the particles traverse the search space adjusting their positions with information from neighbouring particles and its own past positions. Let us consider $x_i^t$ as the position of the particle *i* at time step *t* in the search space, the movement of particles is governed by the following Equation (55).

$$x_i^{t+1} = x_i^t + v_i^{t+1} \quad (55)$$

In our proposed PSO, Particle position is mapped to ***r (t)*** which is an interval number, the Equation (55) can be restated as given in Equations (56) and (57).

$$x_{i(R)}^{t+1} = x_{i(R)}^t + v_{i(R)}^{t+1} \quad (56)$$

$$x_{i(L)}^{t+1} = x_{i(L)}^t + v_{i(L)}^{t+1} \quad (57)$$

$x_{i(R),}^t\, x_{i(L),}^t [x_{min}, x_{max}]$ where $x_{min}, x_{max}$ are [**r**$_R$**, r**$_L$] Interval values tabulated in the Table 5. $v_{i(R)}^t$ , $v_{i(L)}^t$ are the velocities of the ith particle pair in the swarm at time t. The cognitive influence and social influence from all the particles to move towards known better solutions is governed by velocity parameter. $x_{i(R),}^t\, x_{i(L),}^t$ particle pair positions are updated by summing velocities iteratively and used for computing the objective function.

Two variants of PSO, Global best (*Gbest*) PSO and Local best (*Lbest*) PSO are considered for updating the velocities are presented below. In *Gbest* PSO method, best position among all the particles in the swarm influences the position of every particle. The star topology is used to gather the best known solutions from all the particles in the swarm. The *Lbest* PSO method permits every particle's position to be governed by the best known position of particles in the neighbourhood, and it resemble a ring topology. In both the methods, every particle pair has a current position $x_{i(R,L)}^t$, current velocity $v_{i(R,L)}^t$ and a personal best position $P_{best,\, i(R,\, L)}$. The personal best position denotes largest value of a particle pair i from initialization through time t, as determined by the objective function for a maximization problem. The maximum value among all the $P_{best,\, i(R,\, L)}$ is called global best position denoted by $G_{best(R,\, L)}$. In *Lbest* PSO, $L_{best,\, i(R,\, L)}$ is the best solution or best position of any particle pair i that has had from initialization through time t. $P_{best,\, i(R,\, L)}$, $G_{best(R,\, L)}$, $L_{best,\, i(R,\, L)}$ are updated using the Definition 1.3 (mentioned in section III) for optimistic decision making for maximization problems.

The governing equations for the velocities for a particle pair *i* in *Gbest* PSO are given in Equations (58) and (59).

$$v_{i(R)}^{t+1} = \omega v_{i(R)}^t + \varphi_1 r_1^t \left[P_{best,i(R)}^t - x_{i(R)}^t\right] + \varphi_2 r_2^t \left[G_{best(R)} - x_{i(R)}^t\right] \quad (58)$$

$$v_{i(L)}^{t+1} = \omega v_{i(L)}^t + \varphi_1 r_1^t \left[P_{best,i(L)}^t - x_{i(L)}^t\right] + \varphi_2 r_2^t \left[G_{best(L)} - x_{i(L)}^t\right] \quad (59)$$

The governing equations for the velocities for a particle pair i in Lbest PSO methods are given in Equations (60) and (61).

$$v_{i(R)}^{t+1} = \omega v_{i(R)}^t + \varphi_1 r_1^t \left[P_{best,i(R)}^t - x_{i(R)}^t\right] + \varphi_2 r_2^t \left[L_{best,i(R)} - x_{i(R)}^t\right] \quad (60)$$

$$v_{i(L)}^{t+1} = \omega v_{i(L)}^t + \varphi_1 r_1^t \left[P_{best,i(L)}^t - x_{i(L)}^t\right] + \varphi_2 r_2^t \left[L_{best,i(L)} - x_{i(L)}^t\right] \quad (61)$$

Interval arithmetic mentioned in the Definition 1.1 and 1.2 are used in evaluating the Equations (58), (59), (60) and (61). In the Equations (58), (59),(60) and (61), $\boldsymbol{\omega}$ , time varying inertia weight is adopted here, and is found to improve the performance linearly and control the impact of previous velocity on the updated velocity.

Detailed description of all parameters used in the Equations (58), (59),(60) and (61) are presented below.

$G_{best(R,\, L)}$: is the global best position of a particle pair i from initialization through time t;

$L_{best,\, i(R,\, L):}$ is the best position of a particle pair i has had in the neighbourhood from initialization through time t;

$P_{best,\,i(R,L)}$ : is the personal best position for a particle pair *i* from initialization through time t;

$v^t_{i(R,L)}$ : velocity of a particle pair *i* at time t;

$x^t_{i(R,L)}$ : is the position of a particle i at time t;

$\omega$ : is the inertia weight, is equal to $(w_1 - w_2)((T_{max} - t)/T_{max}) + w_2$ , where w1 and w2 are the initial and final weights respectively, t is the current iteration number and $T_{max}$ is the number of iterations.

$\varphi_1$ *and* $\varphi_2$ : positive acceleration co-efficients, also called cognitive and social weights;

$r^t_1$ *and* $r^t_2$ : random numbers from uniform distribution U (0,1) at time t.

Finally, the particle pair positions are updated with velocities for all particle pairs. The objective function is evaluated for every particle pair and using the same, convergence of the solutions is examined. Steps are repeated to compute $P_{best,\,i(R,\,L)}$, $G_{best(R,\,L)}$, $L_{best,\,i(R,\,L)}$ in the next iteration. Stepwise description of proposed Gbest/Lbest PSO algorithm is given below.

## 9.1 Stepwise description of the proposed *Gbest/Lbest PSO* algorithms

The *Gbest/Lbest PSO* algorithm suitably modified for interval RAP is implemented as follows.

1. Particle (pair) population is initialized. Every particle pair is evaluated using the objective function. Initially $P_{best,i(R,L)}$ (Personal best) of a particle is the particle itself. Set of particles pairs are selected from a known size of particles pool. The interval numbers mentioned in the Table 5 are assigned to particle pairs such that it maximizes the system reliability.

2. a. $G_{best(R,L)}$ is evaluated considering all the particle pairs.

2. b. $L_{best,i(R,L)}$ is evaluated from the particles in the neighbourhood.

3. a. Velocities of the particle pairs are computed using Equations (58), (59) *Gbest PSO*.

3. b. Velocities of the particle pairs are computed using Equations (60), (61) in Lbest PSO.

4. Updated velocities for every particle are archived.

5. $P_{best,\,i(R,\,L)}$, $G_{best(R,\,L)}$, $L_{best,\,i(R,\,L)}$ are updated using the Definition 1.3 for optimistic decision making for maximization problems.

6. The iteration counter is incremented. Step 3 and 4 are repeated and the algorithm is terminated when the convergence in the solution occurs.

## 9.2 Common Code fragment outlining the proposed *Gbest PSO* / *Lbest PSO* Algorithms.

1. The particle positions $x^0_{i(R,L)}$, velocities $v^0_{i(R,L)}$, acceleration co-efficients $\varphi_1, \varphi_2$ are initialised. $f^0_{i(R,L)}$ is computed using $x^0_{i(R,L)}$ is computed at time t =0 using the Equation 6.14. The number of particles (P) and iterations (N) are appropriately set.

2. $r^t_1$ *and* $r^t_2$ are stochastically initialized.

*Repeat*

3. a. $v^{t+1}_{i(R)} = \omega v^t_{i(R)} + \varphi_1 r^t_1[P^t_{best,i(R)} - x^t_{i(R)}] + \varphi_2 r^t_2[G_{best(R)} - x^t_{i(R)}]$ and $v^{t+1}_{i(L)} = \omega v^t_{i(L)} + \varphi_1 r^t_1[P^t_{best,i(L)} - x^t_{i(L)}] + \varphi_2 r^t_2[G_{best(L)} - x^t_{i(L)}]$ are used to update velocities in *Gbest* algorithm.

3. b. $v^{t+1}_{i(R)} = \omega v^t_{i(R)} + \varphi_1 r^t_1[P^t_{best,i(R)} - x^t_{i(R)}] + \varphi_2 r^t_2[L_{best,i(R)} - x^t_{i(R)}]$ and $v^{t+1}_{i(L)} = \omega v^t_{i(L)} + \varphi_1 r^t_1[P^t_{best,i(L)} - x^t_{i(L)}] + \varphi_2 r^t_2[L_{best,i(L)} - x^t_{i(L)}]$ are used to update velocities in *Lbest* algorithm

3. c. $r^t_1$ *and* $r^t_2$ are randomly generated for each particle pair for every iteration.

4. $x_{i(R,L)}^{t+1} = x_{i(R,L)}^{t} + v_{i(R,L)}^{t+1}$

*While (i < P)*

5. Compute the objective function $f_{i(R,L)}^{t}$ for every particle $x_{i(R,L)}^{t}$. Convergence in the solution pairs is examined. The steps are repeated until solution converges.

6. If t < N goto step 2.

## 9.3 Evaluation for Particle Swarm Optimization

Two problems are considered for evaluation - evaluation example-one and evaluation example-two as mentioned earlier. In evaluation example-one, six functions and three OICs is considered. In evaluation example-two, 10 functions and six OICs is considered. The initialization and evaluation of model parameters of OSS used in the estimation of system reliability is identical to usage in GA evaluation.

In evaluation example-one , interval RAP for multi-core system with three OICs each performing six functions is considered in the objective function (Equation 47) evaluated using *interval – Lbest* and *interval – Gbest PSO* algorithms. It has a population size of 30 with an archive size of 15 and 50 iterations. The values of $\varphi_1, \varphi_2$ , $w_1$ and $w_2$ have been taken as 0.99876, 0.99678, 0.8999 and 0.2466 respectively. The wakeup probability and readiness parameters are tabulated in Table 6. The cost $C_{ij}$, the execution time in clock cycles for the function j on ith OIC is shown in Table 6. In Equation 45, C is set to 50 clock cycles for the evaluation example one. The optimal solution given by *interval – Lbest* and *interval – Gbest PSO* algorithms are tabulated in Table 12. It is evident from the Table 12 both proposed *interval – Lbest* and *interval – Gbest PSO* algorithms has given better solutions than proposed GA and Regular GA for evaluation example-one problem.

**Table 12 Optimal solution for evaluation example-one with 3OICs and 6 functions using PSO**

| **Description** | **Table 6.1 (set 1)** | **Table 6.1 (set 2)** | **Table 6.1 (set 3)** | **Table 6.1 (set 4)** | **Table 6.1 (set 5)** |
|---|---|---|---|---|---|
| **X** | 101000<br>000010<br>101101 | 110100<br>000111<br>011000 | 010000<br>000011<br>110111 | 111011<br>000110<br>001000 | 111110<br>001111<br>011110 |
| **A** U:<br>OICs | 2 2 2 3 3 3<br>3 | 2 2 2 3 3 3<br>3 | 2 2 2 3 3 3<br>3 | 2 2 2 3 3 3<br>3 | 2 2 3 3 3 2<br>3 |
| ***Interval Gbest PSO*** System Reliability Lower/Upper | 0.970276/ 0.970299 | 0.970281/ 0.970299 | 0.969060/ 0.970377 | 0.970274/ 0.970299 | 0.970276/ 0.970281 |
| ***Interval Lbest PSO*** System Reliability Lower/Upper | 0.969423/ 0.970289 | 0.968900 0.970282 | 0.968960/ 0.970282 | 0.969440/ 0.970283 | 0.970189/ 0.970281 |
| Cost (in cycles) | 44 | 40 | 46 | 49 | 41 |

In evaluation example-two, interval RAP for multi-core system with six OICs each performing ten functions is considered in the objective function (Equation (47)) evaluated using *interval – Lbest* and *interval – Gbest PSO* algorithms. It has a population size of 50 with an archive size of 15 and 100 iterations. The values of $\varphi_1, \varphi_2$, $w_1$ and $w_2$ have been taken as 1.69876, 0.19678, 0.20 and 0.10 respectively. The 10 functions are designated as $F_1$, $F_2$, $F_3$,…, $F_{10}$. The wakeup probability, cost $C_{ij}$ (and C = 3000 cycles) and readiness parameters are tabulated in the Table 9. The optimal solution given by interval – Lbest and interval – Gbest PSO algorithms are tabulated in Table 13 and 14. In both evaluation examples, *Gbest PSO* converges to a better solution than *Lbest PSO*.

Finally, it is known from GA and PSO, MCS - OIC has robust reliability with increase in number of OICs. The conclusion for this paper is presented in the next section.

**Table 13 Optimal solution for evaluation example-two with 6OICs and 10 functions using PSO**

| **Description** | **Table 6.1(set 1)** | **Table 6.1(set 2)** |
|---|---|---|
| **X** | 1 0 0 0 0 0 1 1 1 0<br>0 0 0 0 0 0 1 1 0 0<br>0 0 0 0 0 0 0 0 0 1<br>0 1 0 0 0 0 0 0 0 0<br>0 1 0 0 1 1 0 1 1 0<br>0 0 0 0 1 0 0 0 0 1 | 0 1 0 0 0 1 0 0 1 0<br>0 1 0 0 0 0 1 1 0 0<br>0 0 1 0 1 0 1 0 0 0<br>0 0 0 0 0 0 0 0 0 1<br>1 0 0 0 1 1 0 0 1 0<br>0 0 0 1 0 0 0 0 0 0 |
| U:<br>**A**<br>OICs | 3 3 3 1 3 3 3 3 4 4<br><br>6 | 5 2 4 2 3 2 4 3 3 2<br><br>6 |
| ***Interval Gbest PSO***<br>System Reliability<br>Lower/Upper | 0.999977/0.999997 | 0.999945/0.999984 |
| ***Interval Lbest PSO***<br>System Reliability<br>Lower/Upper | 0.999778/0.999913 | 0.999680/0.999883 |
| Cost (in cycles) | 2500 | 1900 |

**Table 14 Optimal solution for evaluation example-two with 6OICs and 10 functions using PSO**

| Description | Table 6.1(set 3) | Table 6.1(set 4) | Table 6.1(set 5) |
|---|---|---|---|
| **X** | 0 0 0 0 0 0 1 1 0 0<br>0 1 0 0 0 0 0 0 0 0<br>0 0 0 0 0 0 1 1 0 0<br>0 0 0 0 0 0 1 0 0 0<br>0 0 0 1 0 0 0 0 1 0<br>0 0 0 0 0 1 0 1 1 0 | 0 0 0 1 0 1 0 1 0 0<br>0 1 0 0 0 0 0 0 0 0<br>1 0 0 1 0 0 0 0 1 0<br>1 0 0 0 1 0 0 0 0 0<br>0 0 0 0 0 0 0 1 0 0<br>0 0 0 1 1 0 0 1 0 1 | 0 0 0 0 0 0 1 1 0 1<br>0 0 0 0 0 1 1 0 0 0<br>0 1 0 0 0 0 0 0 0 0<br>0 0 1 0 0 1 0 0 0 1<br>0 0 0 0 0 1 0 0 0 1<br>0 0 0 0 1 0 0 0 0 0 |
| **A** U:<br>OICs | 3 3 2 2 3 3 3 2 3 3<br>6 | 4 3 3 3 3 2 3 3 3 4<br>6 | 2 3 3 3 3 3 4 3 3 3<br>6 |
| ***Interval Gbest PSO*** System Reliability Lower/Upper | 0.999984/0.999998 | 0.999955/0.999980 | 0.999989/0.999997 |
| ***Interval Lbest PSO*** System Reliability Lower/Upper | 0.999235/0.999766 | 0.999404/0.999592 | 0.999846/0.999947 |
| Cost (in cycles) | 1900 | 2150 | 2000 |

## 10. Conclusions

(a) The MCS – OIC is a novel economical redundancy-based fault tolerant solution for multi-core system. Power and Area for different configurations of MCS-OICs has been analysed using ASIC simulation. The basic model of MCS-OIC with one MIPS core + one OIC consumes power of 1.1554W and area of 306283µm$^2$. It is observed that maximum power overhead and maximum area overhead is 0.46% and 11.4% respectively, is relatively less when compared to its counterparts. Reliability of the multi-core system consisting of warm standby redundant OICs is modelled and analysed. Actions of OICs are modelled using One Shot System. Model parameters of OSS are mapped to the three different configurations of MCS-OIC. The system reliability and MTTF of MCS-OIC are derived. The system reliability is also derived for four special cases.

(b) System reliabilities for a system consisting of two conventional cores with two, three, four OICs and a system consisting of three conventional cores with three, four, five OICs are studied using the reliability function of the Erlang distribution. The enhancement in the system reliability with increasing number of OICs is observed. That is, for L number of conventional cores with M number of OICs (M > L), a higher system reliability is achieved. Hence, 1: N model provides a scalable design alternative.

(c) A two-phase GA is used to solve Interval RAP. Optimal solutions for evaluation example-one and evaluation example-two are obtained. Results indicate that the proposed GA evolves better solution than regular GA. System reliability monotonically increasing with respect to function and component level redundancy is proved. It is an inherent characteristic of the proposed Interval RAP. It is understood that this property ensures enhancement of the system reliability for multi-core systems with OICs.

(d) Results of sensitivity analysis indicate stability in the system. GA parameters are fixed resulting in better solution for maximizing system reliability using sensitivity analysis. It is observed that variations in GA parameters do not affect system stability and the optimal solution.

(e) Two variants of PSO, Global best (Gbest) PSO and Local best (Lbest) are used to solve Interval RAP. For evaluation example-one, results indicate proposed GBest and Lbest provide better results compared to proposed GA. In evaluation example-two, insignificant difference in solutions given by PSO variants and GA is observed. Results given by PSO also indicate system reliability increasing with respect to function and component level redundancy.

(f) Considering GA and PSO evaluation of interval RAP, It is inferred that the system is stable and optimal solution does not change due to interval numbers, and enhancement in the system reliability is ensured by the monotone property of the Interval RAP.

## References

The citations to research articles appear in the order of occurrence in the manuscript.